\RequirePackage{fix-cm}
\documentclass[twocolumn]{svjour3}          
\smartqed  
\usepackage{subfig}
\usepackage{graphicx}
\usepackage{natbib}
\usepackage{amsmath,amsfonts,amssymb}
\usepackage{bbm}
\usepackage[colorlinks,citecolor=blue,urlcolor=blue]{hyperref}
\usepackage{booktabs}
\usepackage{float}
\usepackage{tikz}
\usetikzlibrary{fit,positioning,calc}
\usepackage{scrextend}
\usepackage[switch]{lineno}
\usepackage[ruled,norelsize,vlined]{algorithm2e}
\usepackage{enumitem}
\usepackage[T1]{fontenc}
\usepackage[utf8]{inputenc}
        \usepackage{setspace}

\SetAlFnt{\small}
\SetAlCapFnt{\small}
\SetAlCapNameFnt{\small}

\newcommand{\br}{ {\bf r} }

\newcommand{\bP}{ {\bf P} }
\newcommand{\bS}{ {\bf S} }
\newcommand{\bs}{ {\bf s} }
\newcommand{\bU}{ {\bf U} }
\newcommand{\by}{ {\bf y} }

\newcommand{\bu}{ {\bf u} }

\newcommand{\bD}{ {\bf D} }
\newcommand{\bI}{ {\bf I} }

\newcommand{\bV}{ {\bf V} }
\newcommand{\bF}{ {\bf F} }
\newcommand{\bG}{ {\bf G} }
\newcommand{\bW}{ {\bf W} }

\newcommand{\bB}{ {\bf B} }

\newcommand{\ba}{ {\bf a} }
\newcommand{\bc}{ {\bf c} }

\newcommand{\bgamma}{ {\boldsymbol \gamma} }

\newcommand{\bvarepsilon}{ {\boldsymbol \varepsilon} }
\newcommand{\bLambda}{ {\boldsymbol \Lambda} }

\newcommand{\bGamma}{ {\boldsymbol \Gamma} }
\newcommand{\bDelta}{ {\boldsymbol \Delta} }

\newcommand{\bmu}{ {\boldsymbol \mu} }

\newcommand{\bSigma}{ {\boldsymbol \Sigma} }

\newcommand{\bOmega}{ {\boldsymbol \Omega} }
\newcommand{\bomega}{ {\boldsymbol \omega} }

\newcommand{\bxi}{ {\boldsymbol \xi} }
\newcommand{\bz}{ {\bf z} }
\newcommand{\btheta}{ {\boldsymbol \theta} }

\newtheorem{Lemma}{Lemma}
\newtheorem{Theorem}{Theorem}
\newtheorem{Proposition}{Proposition}
\newtheorem{Corollary}{Corollary}
\newtheorem{Remark}{Remark}

\begin{document}
\fontfamily{lmss}\selectfont
	\setstretch{1.26}

\title{\fontfamily{lmss}\selectfont
 A closed-form filter for binary time series
}

\author{\fontfamily{lmss}\selectfont Augusto Fasano\textsuperscript{1} \hspace*{-5.4pt} \and
        \hspace*{-5.4pt} Giovanni Rebaudo\textsuperscript{2} \hspace*{-5.4pt} \and
        \hspace*{-5.4pt} Daniele Durante\textsuperscript{3} \hspace*{-5.4pt} \and
        \hspace*{-5.4pt} Sonia Petrone\textsuperscript{3}
}

\twocolumn[
\maketitle
\begin{@twocolumnfalse}
\vspace{-84pt}
\begin{abstract}
Non-Gaussian state-space models arise in several applications, and within this framework the binary time series setting provides a relevant example. However, unlike for Gaussian state-space models --- where filtering, predictive and smoothing distributions are available in closed form --- binary state-space models require  approximations or sequential Monte Carlo strategies for inference and prediction. This is due to the apparent absence of conjugacy between the Gaussian states and the likelihood induced by the observation equation for the binary data. In this article we prove that the filtering, predictive and smoothing distributions in dynamic probit models with Gaussian state variables are, in fact, available and belong to a class of unified skew-normals (\textsc{sun}) whose parameters can be updated recursively  in time via analytical expressions. Also the key functionals of these distributions are, in principle, available, but their calculation requires the evaluation of multivariate Gaussian cumulative distribution functions. Leveraging \textsc{sun} properties, we address this issue via novel Monte Carlo methods based on independent  samples from the smoothing distribution, that can easily be adapted to the filtering and predictive case, thus improving state-of-the-art approximate and sequential Monte Carlo inference in small-to-moderate dimensional studies. Novel sequential Monte Carlo procedures that exploit the \textsc{sun} properties are also developed to deal with  online inference in high dimensions. Performance gains over competitors are outlined in a financial application.

\end{abstract}	
\vspace{20pt}
\end{@twocolumnfalse}
]

{
	\renewcommand{\thefootnote}%
	{}
	\footnotetext[1]{\fontfamily{lmss}\selectfont Corresponding author: Augusto Fasano\\
	\hspace*{12pt} augusto.fasano@unito.it}
}
{
	\renewcommand{\thefootnote}%
	{\arabic{footnote}}
	\footnotetext[1]{\fontfamily{lmss}\selectfont ESOMAS Department, University of Turin, and Collegio\\ \hspace*{12pt} Carlo Alberto, Turin, Italy}
}
{
	\renewcommand{\thefootnote}%
	{\arabic{footnote}}
	\footnotetext[2]{\fontfamily{lmss}\selectfont Department of Statistics and Data Sciences, the Univer-\\ \hspace*{12pt} sity of Texas at Austin, Austin, United States of America}
}
{
	\renewcommand{\thefootnote}%
	{\arabic{footnote}}
	\footnotetext[3]{\fontfamily{lmss}\selectfont Department of Decision Sciences and Institute for Data\\ \hspace*{12pt} Science and Analytics, Bocconi University, Milan, Italy}
}

\section{Introduction}\label{sec_1}

Despite the availability of several alternative approaches for dynamic inference and prediction of binary time series \citep{macdonald1997}, state-space models are a source of constant interest due to their flexibility in accommodating a variety of representations and dependence structures via an interpretable formulation \citep{west2006, petris2009, durbin2012}. Let $\by_{t}=(y_{1t}, \ldots, y_{mt})^{\intercal} \in \{0;1\}^m$ be a vector of binary event data observed at time $t$, and denote with $\btheta_{t}=(\theta_{1t}, \ldots, \theta_{pt})^{\intercal} \in \mathbb{R}^p$ the corresponding vector of state variables. Adapting the notation in, e.g., \citet{petris2009} to our setting, we aim to provide closed-form expressions for the filtering, predictive and smoothing distributions in the general multivariate dynamic probit model 
\begin{align}
&p(\by_{t} \mid \btheta_{t})= \Phi_m(\bB_t\bF_t\btheta_{t}; \bB_t\bV_t\bB_t), \label{eq1}\\ 
&\btheta_t=\bG_{t}\btheta_{t-1}+\bvarepsilon_t, \quad \bvarepsilon_t \sim  \mbox{N}_p({\bf 0}, \bW_t), \ t=1 \ldots, n, \label{eq2}
\end{align}
with $\btheta_0 \sim \mbox{N}_p(\ba_0, \bP_0)$, and dependence structure as defined by the directed acyclic graph displayed in Fig.~\ref{F_mod1}. In \eqref{eq1}, $ \Phi_m(\bB_t\bF_t\btheta_{t}; \bB_t\bV_t\bB_t)$ is the cumulative distribution function of a $\mbox{N}_m({\bf 0},\bB_t\bV_t\bB_t)$ evaluated at $\bB_t\bF_t\btheta_{t}$, with $\bB_t=\mbox{diag}(2y_{1t}-1, \ldots,  2y_{mt}-1)$ denoting the $m \times m$ sign matrix associated with $\by_t$, which defines the multivariate probit likelihood in \eqref{eq1}.

\begin{figure*}[t]
	\centering
	\begin{tikzpicture}
	\tikzstyle{main}=[minimum size = 14mm, thick, draw =black!80, node distance = 6mm]
	\tikzstyle{connect}=[-latex, thick]
	\tikzstyle{box}=[rectangle, draw=black!100]
	
	\node[main, circle, fill = white!100] (theta0) {$\btheta_0$ };
	\node[main, circle, fill = white!100] (theta1) [right=of theta0]{$\btheta_1$ };
	\node[main, circle, fill = white!100] (theta2) [right=of theta1]{$\btheta_2$ };
	\node[main, circle, fill = white!100,draw=white] (dots1) [right=of theta2] {$\cdots$ };
	\node[main, circle, fill = white!100] (thetat) [right=of dots1] {$\btheta_t$};
	\node[main, circle, fill = white!100,draw=white] (dots2) [right=of thetat] {$\cdots$ };
	\node[main, circle, fill = white!100] (thetan1) [right=of dots2] {$\btheta_{n-1}$};
	\node[main, circle, fill = white!100] (thetan) [right=of thetan1] {$\btheta_n$};
	
	\node[main, circle, fill = white!100, dotted]  (epsilon1) [above=of theta1] {$\bvarepsilon_1$ };
	\node[main, circle, fill = white!100, dotted]  (epsilon2) [above=of theta2] {$\bvarepsilon_2$ };
	\node[main, circle, fill = white!100, draw=white] (dots11) [above=of dots1] {$\cdots$ };
	\node[main, circle, fill = white!100, dotted]  (epsilont) [above=of thetat] {$\bvarepsilon_t$ };
	\node[main, circle, fill = white!100, draw=white] (dots12) [above=of dots2] {$\cdots$ };
	\node[main, circle, fill = white!10, dotted]  (epsilonn1) [above=of thetan1] {$\bvarepsilon_{n-1}$ };
	\node[main, circle, fill = white!100, dotted]  (epsilonn) [above=of thetan] {$\bvarepsilon_n$ };
		
	\node[main, fill = black!10] (y1) [below=of theta1] {$\by_1$ };
	\node[main, fill = black!10] (y2) [below=of theta2] {$\by_2$ };
	\node[main, fill = white!10,draw=white] (dots11) [below=of dots1] {$\cdots$ };
	\node[main, fill = black!10] (yt) [below=of thetat] {$\by_t$ };
	\node[main, fill = white!10,draw=white] (dots12) [below=of dots2] {$\cdots$ };
	\node[main, fill = black!10] (yn1) [below=of thetan1] {$\by_{n-1}$ };
	\node[main, fill = black!10] (yn) [below=of thetan] {$\by_n$ };
	
	\path        (epsilon1) edge [connect] (theta1);
	\path        (epsilon2) edge [connect] (theta2);
	\path    (epsilont) edge [connect] (thetat);
	\path      (epsilonn1) edge [connect] (thetan1);
	\path      (epsilonn) edge [connect] (thetan);
	\path        (theta1) edge [connect] (y1);
	\path        (theta2) edge [connect] (y2);
	\path    (thetat) edge [connect] (yt);
	\path      (thetan1) edge [connect] (yn1);
	\path      (thetan) edge [connect] (yn);
	\path        (theta0) edge [connect] (theta1);
	\path        (theta1) edge [connect] (theta2);
	\path        (theta2) edge [connect] (dots1);
	\path        (dots1) edge [connect] (thetat);
	\path        (thetat) edge [connect] (dots2);
	\path        (dots2) edge [connect] (thetan1);
	\path        (thetan1) edge [connect] (thetan);
	\end{tikzpicture}
	\caption{\footnotesize{\fontfamily{lmss}\selectfont Graphical representation of model \eqref{eq1}--\eqref{eq2}. The dashed circles, solid circles and grey squares denote Gaussian errors, Gaussian states and observed binary data, respectively.}}\label{F_mod1}
\end{figure*}

\begin{figure*}[t]
	\centering
	\begin{tikzpicture}
	\tikzstyle{main}=[minimum size = 14mm, thick, draw =black!80, node distance = 6mm]
	\tikzstyle{connect}=[-latex, thick]
	\tikzstyle{box}=[rectangle, draw=black!100]
	
	\node[main, circle, fill = white!100] (theta0) {$\btheta_0$ };
	\node[main, circle, fill = white!100] (theta1) [right=of theta0]{$\btheta_1$ };
	\node[main, circle, fill = white!100] (theta2) [right=of theta1]{$\btheta_2$ };
	\node[main, circle, fill = white!100,draw=white] (dots1) [right=of theta2] {$\cdots$ };
	\node[main, circle, fill = white!100] (thetat) [right=of dots1] {$\btheta_t$};
	\node[main, circle, fill = white!100,draw=white] (dots2) [right=of thetat] {$\cdots$ };
	\node[main, circle, fill = white!100] (thetan1) [right=of dots2] {$\btheta_{n-1}$};
	\node[main, circle, fill = white!100] (thetan) [right=of thetan1] {$\btheta_n$};
	
	\node[main, circle, fill = white!100, dotted]  (epsilon1) [above=of theta1] {$\bvarepsilon_1$ };
	\node[main, circle, fill = white!100, dotted]  (epsilon2) [above=of theta2] {$\bvarepsilon_2$ };
	\node[main, circle, fill = white!100, draw=white] (dots11) [above=of dots1] {$\cdots$ };
	\node[main, circle, fill = white!100, dotted]  (epsilont) [above=of thetat] {$\bvarepsilon_t$ };
	\node[main, circle, fill = white!100, draw=white] (dots12) [above=of dots2] {$\cdots$ };
	\node[main, circle, fill = white!10, dotted]  (epsilonn1) [above=of thetan1] {$\bvarepsilon_{n-1}$ };
	\node[main, circle, fill = white!100, dotted]  (epsilonn) [above=of thetan] {$\bvarepsilon_n$ };
		
	\node[main, fill = white!10] (z1) [below=of theta1] {$\bz_1$ };
	\node[main, fill = white!10] (z2) [below=of theta2] {$\bz_2$ };
	\node[main, fill = white!10,draw=white] (dots11) [below=of dots1] {$\cdots$ };
	\node[main, fill = white!10] (zt) [below=of thetat] {$\bz_t$ };
	\node[main, fill = white!10,draw=white] (dots12) [below=of dots2] {$\cdots$ };
	\node[main, fill = white!10] (zn1) [below=of thetan1] {$\bz_{n-1}$ };
	\node[main, fill = white!10] (zn) [below=of thetan] {$\bz_n$ };
	
	\node[main, fill = black!10] (y1) [below=of z1] {$\by_1$ };
	\node[main, fill = black!10] (y2) [below=of z2] {$\by_2$ };
	\node[main, fill = white!10,draw=white] (dots21) [below=of dots11] {$\cdots$ };
	\node[main, fill = black!10] (yt) [below=of zt] {$\by_t$ };
	\node[main, fill = white!10,draw=white] (dots22) [below=of dots12] {$\cdots$ };
	\node[main, fill = black!10] (yn1) [below=of zn1] {$\by_{n-1}$ };
	\node[main, fill = black!10] (yn) [below=of zn] {$\by_n$ };
	
	\path        (epsilon1) edge [connect] (theta1);
	\path        (epsilon2) edge [connect] (theta2);
	\path    (epsilont) edge [connect] (thetat);
	\path      (epsilonn1) edge [connect] (thetan1);
	\path      (epsilonn) edge [connect] (thetan);
	\path        (theta1) edge [connect] (z1);
	\path        (theta2) edge [connect] (z2);
	\path    (thetat) edge [connect] (zt);
	\path      (thetan1) edge [connect] (zn1);
	\path      (thetan) edge [connect] (zn);
	\path        (z1) edge [connect] (y1);
	\path        (z2) edge [connect] (y2);
	\path    (zt) edge [connect] (yt);
	\path      (zn1) edge [connect] (yn1);
	\path      (zn) edge [connect] (yn);
	\path        (theta0) edge [connect] (theta1);
	\path        (theta1) edge [connect] (theta2);
	\path        (theta2) edge [connect] (dots1);
	\path        (dots1) edge [connect] (thetat);
	\path        (thetat) edge [connect] (dots2);
	\path        (dots2) edge [connect] (thetan1);
	\path        (thetan1) edge [connect] (thetan);
	\end{tikzpicture}
	\caption{\footnotesize{\fontfamily{lmss}\selectfont Graphical representation of model \eqref{eq3}--\eqref{eq5}.  Dashed circles, solid circles, white squares and grey squares denote Gaussian errors, Gaussian states, latent Gaussian data and observed binary data, respectively.}}\label{F_mod}
\end{figure*}

Model \eqref{eq1}--\eqref{eq2} generalizes  univariate dynamic probit models to multivariate settings, as we will clarify in equations \eqref{eq3}--\eqref{eq5}. The quantities $\bF_t,\bV_t, \bG_t, \bW_t, \ba_0$ and $\bP_0$ denote, instead, known matrices controlling the location, scale and dependence structure in the state-space model \eqref{eq1}--\eqref{eq2}. Estimation and inference for these matrices is, itself, a relevant problem which can be addressed  both from a frequentist and a Bayesian perspective. Yet our focus  is on providing exact results for inference on state variables  and  prediction of future binary events under  \eqref{eq1}--\eqref{eq2}. Therefore, consistent with the classical Kalman filter \citep{kalman1960}, we rely on known system matrices $\bF_t,\bV_t, \bG_t, \bW_t, \ba_0$ and $\bP_0$. Nonetheless, novel results on marginal likelihoods, which can be used in parameter estimation, are provided in Sect.~\ref{subsec:Smoothing}.

Model \eqref{eq1}--\eqref{eq2} provides a general representation encompassing a variety of formulations. For example, setting $\bV_t=\bI_m$ in \eqref{eq1}  for each $t$ yields a set of standard dynamic probit regressions, which include  the classical univariate dynamic probit model when $m=1$. These representations have appeared in several applications, especially within the econometrics literature, due to a direct connection between  \eqref{eq1}--\eqref{eq2} and dynamic discrete choice models \citep{keane2009}.  This is due to the fact that representation \eqref{eq1}--\eqref{eq2} can be alternatively obtained via the dichotomization of an underlying state-space model for the $m$-variate Gaussian time series $\bz_t=(z_{1t}, \ldots, z_{mt})^{\intercal}\in \mathbb{R}^m$,  $t=1,\ldots, n$, which is regarded, in  econometric applications, as a set of time-varying utilities. Indeed, adapting classical results from static probit regression \citep{Albert_1993,chib1998}, model  \eqref{eq1}--\eqref{eq2} is equivalent to
\vspace{-2pt}
\begin{equation}
\begin{split}
\by_{t}&=(y_{1t}, \ldots, y_{mt})^{\intercal}= {\boldsymbol{\mathbbm{1}}}(\bz_t > {\bf 0})\\
&=[\mathbbm{1}(z_{1t}>0), \ldots, \mathbbm{1}(z_{mt}>0)]^{\intercal}, \quad t=1, \ldots, n,
\end{split}
\label{eq3}
\end{equation}
with $\bz_1, \ldots, \bz_n$ evolving in time according to the Gaussian state-space model
\vspace{-2pt}
\begin{align}
&p(\bz_{t} \mid \btheta_t)= \phi_m(\bz_{t}-\bF_t\btheta_{t};\bV_t), \label{eq4}\\ 
&\btheta_t=\bG_{t}\btheta_{t-1}+\bvarepsilon_t, \quad \bvarepsilon_t \sim \mbox{N}_p({\bf 0}, \bW_t), \ t=1 \ldots, n, \label{eq5}
\end{align}
having $\btheta_0 \sim \mbox{N}_p(\ba_0, \bP_0)$ and dependence structure as defined by the directed acyclic graph displayed in Fig.~\ref{F_mod}. In \eqref{eq4}, $\phi_m(\bz_{t}- \bF_t\btheta_{t}; \bV_t)$ denotes the density function of the Gaussian $\mbox{N}_m(\bF_t\btheta_{t},\bV_t)$ evaluated at $\bz_t\in \mathbb{R}^m$. To clarify the connection between \eqref{eq1}--\eqref{eq2} and \eqref{eq3}--\eqref{eq5}, note that if $\tilde{\bz}_t$ is a generic Gaussian random variable with density \eqref{eq4}, then it holds $p(\by_{t} \mid \btheta_t)=\mbox{pr}(\bB_t \tilde{\bz}_t >{\bf 0})=\mbox{pr}[-\bB_t (\tilde{\bz}_t - \bF_t\btheta_{t})<  \bB_t\bF_t\btheta_{t}]=\Phi_m(\bB_t\bF_t\btheta_{t}{;}  \bB_t\bV_t \bB_t)$, given that $-\bB_t (\tilde{\bz}_t - \bF_t\btheta_{t}) \sim \mbox{N}_m({\bf 0}{,}  \bB_t\bV_t \bB_t)$ under  \eqref{eq4}.

As is clear from model  \eqref{eq4}--\eqref{eq5}, if $\bz_{1:t}=(\bz^{\intercal}_1, \ldots, \bz^{\intercal}_t)^{\intercal}$ were observed, dynamic inference on the states $\btheta_t$, for $t=1, \ldots, n$, would be possible via direct application of the  Kalman filter \citep{kalman1960}. Indeed, exploiting Gaussian-Gaussian conjugacy and the conditional independence properties that are represented in Fig.~\ref{F_mod}, the filtering $p(\btheta_t \mid \bz_{1:t})$  and predictive $p(\btheta_t \mid \bz_{1:t-1})$ densities are also Gaussian and have parameters which can be computed recursively via simple expressions relying on the previous updates. Moreover,  the smoothing density $p(\btheta_{1:n} \mid \bz_{1:n})$ and its marginals $p(\btheta_{t} \mid \bz_{1:n})$, $t \leq n$, can also be obtained in closed form leveraging Gaussian-Gaussian conjugacy. However, in \eqref{eq3}--\eqref{eq5} only a dichotomized version $\by_t$ of $\bz_t$ is available. Therefore, the filtering, predictive and smoothing densities of interest are $p(\btheta_t \mid \by_{1:t})$, $p(\btheta_t \mid \by_{1:t-1})$ and $p(\btheta_{1:n} \mid \by_{1:n})$, respectively. Recalling model \eqref{eq1}--\eqref{eq2} and Bayes' rule, the calculation of these quantities proceeds by updating the Gaussian distribution for the states in \eqref{eq2} with the  probit likelihood in  \eqref{eq1}, thereby providing conditional distributions which do not have an obvious closed form \citep{Albert_1993,chib1998}.

When the focus is on online inference for filtering and prediction, one solution to the above issue is to rely on approximations of model \eqref{eq1}--\eqref{eq2} which allow the implementation of standard Kalman filter updates, thus leading to approximate dynamic inference on the states via extended \citep{uhlmann1992} or unscented \citep{julier1997} Kalman filters, among others. However, these approximations may lead to unreliable  inference in various settings \citep{andrieu2002}. Markov chain Monte Carlo (\textsc{mcmc})  strategies \citep[e.g.,][]{carlin1992,shephard1994, soyer2013bayesian} address this problem but, unlike the Kalman filter, these methods are only suitable for batch learning of smoothing distributions, and  tend to face mixing or scalability issues in binary settings \citep{Johndrow2018}.

Sequential Monte Carlo methods \citep[e.g.,][]{doucet2001} partially solve \textsc{mcmc} issues, and are specifically developed for online inference via particle-based representations of the states' conditional  distributions, which are then propagated in time for dynamic filtering and prediction \citep{gordon1993, kitagawa1996,liu1998,pitt1999, doucet2000,andrieu2002}. These strategies provide state-of-the-art  solutions  in non-Gaussian state-space models, and can be also adapted to perform batch learning of the smoothing distribution; see \citet{Doucet2009} for  a discussion on particles' degeneracy issues that may arise in such a setting. Nonetheless, sequential Monte Carlo is clearly still sub-optimal compared to the case in which $p(\btheta_t \mid \by_{1:t})$, $p(\btheta_t \mid \by_{1:t-1})$ and $p(\btheta_{1:n} \mid \by_{1:n})$ are available in closed form and belong to a tractable class of known densities whose parameters can be sequentially updated via analytical expressions.

In Sect.~\ref{sec.3}, we prove that, for the dynamic multivariate probit model in  \eqref{eq1}--\eqref{eq2}, the quantities $p(\btheta_t \mid \by_{1:t})$, $p(\btheta_t \mid \by_{1:t-1})$ and  $p(\btheta_{1:n} \mid \by_{1:n})$ are  unified skew-normal (\textsc{sun}) densities  \citep{arellano_2006} having tractable expressions for the recursive computation of the corresponding parameters. To the best of our knowledge, such a  result provides the first closed-form filter and smoother for binary time series, and facilitates  improvements both in online and batch inference. As we will highlight in Sect.~\ref{sec.2}, the \textsc{sun} distribution has several closure properties  \citep{arellano_2006, azzalini_2013} in addition to explicit formulas --- involving the cumulative distribution function of multivariate Gaussians --- for the moments \citep{azzalini_2010,gupta_2013} and the normalizing constant \citep{arellano_2006}. In Sect.~\ref{sec.3}, we exploit these properties to derive closed-form expressions for functionals of $p(\btheta_t \mid \by_{1:t})$, $p(\btheta_t \mid \by_{1:t-1})$ and  $p(\btheta_{1:n} \mid \by_{1:n})$, including, in particular, the observations' predictive density $p( \by_{t} \mid  \by_{1:t-1})$ and the marginal likelihood $p(\by_{1:n})$. In Sect.~\ref{sec.41}, we also derive an exact Monte Carlo scheme to compute generic functionals of the smoothing distribution. This routine relies on a generative representation of the \textsc{sun} via linear combinations of multivariate Gaussians and truncated normals \citep{arellano_2006}, and can be also applied effectively  to evaluate  the functionals of filtering and predictive densities in small-to-moderate dimensions where $mt$ is of the order of few hundreds, a common situation in routine applications.

As clarified in Sect.~\ref{sec.42}, the above strategies face computational bottlenecks in higher dimensions \citep{botev_2017}, due to challenges in computing cumulative distribution functions of multivariate Gaussians, and in sampling from multivariate truncated normals. In these contexts, we develop new sequential Monte Carlo methods that exploit  \textsc{sun} properties. In particular, we first prove in Sect.~\ref{sec.421} that an optimal particle filter, in the sense of \citet{doucet2000}, can be derived analytically, thus covering a gap in the literature. This strategy is further improved in  Sect.~\ref{sec.422} via a class of partially collapsed sequential Monte Carlo methods that recursively update via lookahead strategies  \citep{lin2013} the multivariate truncated normal component in the  \textsc{sun} generative additive representation, while keeping the Gaussian part exact. As outlined in an illustrative financial application in Sect.~\ref{sec.5}, this class improves approximation accuracy relative to competing methods, and includes, as a special case, the Rao--Blackwellized particle filter of \citet{andrieu2002}. Concluding remarks can be found in Sect.~\ref{sec.6}.

\section{The unified skew-normal distribution}
\label{sec.2}
\vspace{-5pt}
Before deriving  filtering, predictive and smoothing distributions under model \eqref{eq1}--\eqref{eq2}, let us first briefly review the \textsc{sun} family. \citet{arellano_2006} proposed this broad class to unify different extensions \citep[e.g.,][]{arnold_2000,arnold_2002,gupta_2004,gonz_2004} of the original multivariate skew-normal \citep{azza_1996}, whose density is obtained as the product between a multivariate Gaussian density and the cumulative distribution function of a standard normal evaluated at a value which depends on a skewness-inducing vector of parameters. Motivated by the success of this formulation and of its  generalizations \citep{azza_1999},  \citet{arellano_2006} developed a unifying representation, namely the \textsc{sun} distribution. A random vector $\btheta \in \mathbb{R}^q$ has unified skew-normal distribution, $\btheta \sim \mbox{\textsc{sun}}_{q,h}(\bxi,\bOmega,\bDelta,\bgamma,\bGamma)$, if its density function $p(\btheta)$ can be expressed as
\begin{equation}
\phi_q(\btheta -\bxi; \bOmega) \frac{\Phi_h[\bgamma+\bDelta^\intercal \bar{\bOmega}^{-1} \bomega^{-1}(\btheta-\bxi){;} \bGamma{-}\bDelta^{\intercal}\bar{\bOmega}^{-1}\bDelta ]}{\Phi_h(\bgamma;\bGamma)},
\label{eq6}
\end{equation}
where the covariance matrix $\bOmega$ of the Gaussian density $\phi_q(\btheta -\bxi; \bOmega)$ can be decomposed as $\bOmega=\bomega \bar{\bOmega} \bomega$, that is by re-scaling the $q \times q$ correlation matrix $\bar{\bOmega}$ via the positive diagonal scale matrix $ \bomega=(\bOmega \odot {\bf I}_q)^{1/2}$, with $\odot$ denoting the element-wise Hadamard product. In \eqref{eq6}, the skewness-inducing mechanism is driven by the cumulative distribution function of the $\mbox{N}_h(\boldsymbol{0}, \bGamma-\bDelta^{\intercal}\bar{\bOmega}\ ^{-1}\bDelta)$ computed at $\bgamma+\bDelta^\intercal \bar{\bOmega}\ ^{-1} \bomega^{-1}(\btheta-\bxi)$, whereas $\Phi_h(\bgamma;\bGamma)$ denotes the normalizing constant obtained by evaluating the  cumulative distribution function of a $\mbox{N}_h(\boldsymbol{0}, \bGamma)$ at $\bgamma$. \citet{arellano_2006} added a further identifiability condition which restricts the matrix $\bOmega^*$, with blocks $\bOmega_{[11]}^*=\bGamma$, $\bOmega_{[22]}^*=\bar{\bOmega}$ and $\bOmega_{[21]}^*=\bOmega_{[12]}^{*\intercal}=\bDelta$, to be a full--rank correlation matrix. Note that in \eqref{eq6} the quantities $q$ and $h$ define the dimensions of the Gaussian density and cumulative distribution function, respectively. As clarified by our closed-form \textsc{sun} results in Sect.~\ref{sec.3}, $q$ defines the dimension of the states' vector, and coincides with $p$ in the \textsc{sun} filtering and predictive distributions, while it is equal to $pn$ in the \textsc{sun} smoothing distribution. On the other hand, $h$ increases linearly with time  in all the distributions of interest.

To clarify the role of the parameters in  \eqref{eq6}, we first discuss a stochastic representation of the \textsc{sun}. Let $ \tilde{\bz} \in \mathbb{R}^h$ and $\tilde{\btheta} \in \mathbb{R}^q$ characterize two random vectors jointly distributed as a $\mbox{N}_{h+q}({\bf 0},  \bOmega^*)$, then $(\bxi+\bomega\tilde{\btheta} \mid \tilde{\bz}+\bgamma > {\bf 0}) \sim \mbox{\textsc{sun}}_{q,h}(\bxi,\bOmega,\bDelta,\bgamma,\bGamma)$ \citep{arellano_2006}. Hence, $\bxi$ and $\bomega$ control location and scale, respectively, while $\bGamma$, $\bar{\bOmega}$ and $\bDelta$ define the dependence within  $\tilde{\bz} \in \mathbb{R}^h$, $\tilde{\btheta} \in \mathbb{R}^q$ and between these two vectors, respectively. Finally, $\bgamma$ controls the truncation in the partially observed Gaussian vector $\tilde{\bz}\in \mathbb{R}^h$. The above result provides also relevant insights on our closed-form filter for the dynamic probit model \eqref{eq1}--\eqref{eq2}, which will be further clarified in Sect.~\ref{sec.3}. Indeed, according to  \eqref{eq3}--\eqref{eq5}, the filtering, predictive and smoothing densities induced by model \eqref{eq1}--\eqref{eq2} can be also defined as $p(\btheta_t \mid \by_{1:t})=p[\btheta_t \mid {\boldsymbol{\mathbbm{1}}}(\bz_{1:t}>{\bf 0})]$, $p(\btheta_t \mid \by_{1:t-1})=p[\btheta_t \mid {\boldsymbol{\mathbbm{1}}}(\bz_{1:t-1}>{\bf 0})]$ and $p(\btheta_{1:n} \mid \by_{1:n})=p[\btheta_{1:n} \mid {\boldsymbol{\mathbbm{1}}}(\bz_{1:n}>{\bf 0})]$, respectively, with $(\bz_t, \btheta_t)$ from the Gaussian state-space model  \eqref{eq4}--\eqref{eq5} for $t=1, \ldots, n$, thus highlighting the direct connection between these densities and the stochastic representation of the \textsc{sun}.

An additional generative additive representation of the  \textsc{sun} relies on linear combinations of Gaussian and truncated normal random variables, thereby facilitating sampling from the \textsc{sun}. In particular, recalling \citet[][Sect.~7.1.2]{azzalini_2013} and  \citet{arellano_2006}, if $\btheta \sim \mbox{\textsc{sun}}_{q,h}(\bxi,\bOmega,\bDelta,\bgamma{,}\bGamma)$, then
\begin{equation}
\btheta\stackrel{\mbox{\scriptsize d}}{=}\bxi+\bomega(\bU_0+\bDelta \bGamma^{-1} \bU_1), \quad \bU_0 \perp \bU_1,
\label{eq7}
\end{equation}
with $\bU_0 \sim \mbox{N}_q({\bf 0}{,}\bar{\bOmega}- \bDelta\bGamma^{-1}\bDelta^{\intercal})$ and $\bU_1$ from a $\mbox{N}_h({\bf 0}{,}\bGamma)$ truncated below $-\bgamma$. As clarified in Sect.~\ref{sec.4}, this result can facilitate efficient Monte Carlo inference on complex functionals of \textsc{sun} filtering, predictive and smoothing distributions under model \eqref{eq1}--\eqref{eq2}, leveraging  independent and identically distributed samples from such variables. Indeed, although key moments can be explicitly derived via the  differentiation of the \textsc{sun} moment generating function \citep{gupta_2013,arellano_2006}, such a strategy requires tedious calculations when the focus is on complex functionals. Moreover, recalling  \citet{azzalini_2010} and \citet{gupta_2013}, the first and second order moments further require the evaluation of $h$-variate Gaussian cumulative distribution functions $\Phi_h(\cdot)$, thus affecting computational tractability in large $h$ settings \citep[e.g.,][]{botev_2017}.  In these situations, Monte Carlo integration provides an effective solution, especially when independent samples can be generated efficiently.  Therefore, we mostly focus on improved Monte Carlo inference under model \eqref{eq1}--\eqref{eq2} exploiting the  \textsc{sun} properties, and refer to \citet{azzalini_2010} and \citet{gupta_2013} for a closed-form expression of the expectation, variance and cumulative distribution function of \textsc{sun} variables.

Before concluding this general overview, we  emphasize that \textsc{sun} variables are also closed under marginalization, linear combinations and conditioning \citep{azzalini_2013}. These properties facilitate the derivation of the \textsc{sun} filtering, predictive and smoothing distributions under model \eqref{eq1}--\eqref{eq2}.

\vspace{-8pt}

\section{Filtering, prediction and smoothing}
\label{sec.3}
\vspace{-3pt}
In Sects.~\ref{sec.31} and \ref{subsec:Smoothing}, we prove that all the distributions of  direct interest admit a closed-form \textsc{sun} representation. Specifically, in Sect.~\ref{sec.31} we show that closed-form filters --- meant here  as exact updating schemes for predictive and filtering distributions based on simple recursive expressions for the associated parameters --- can be obtained under model \eqref{eq1}--\eqref{eq2}. Similarly,  in Sect.~\ref{subsec:Smoothing} we derive the form of the \textsc{sun} smoothing distribution and present important consequences. The associated computational methods are then discussed in Sect.~\ref{sec.4}.

\vspace{-10pt}
\subsection{Filtering and predictive distributions}
\label{sec.31}
\vspace{-3pt}
To obtain the exact form of the filtering and predictive distributions under \eqref{eq1}--\eqref{eq2}, let us start from $p(\btheta_1 \mid \by_1)$. This first quantity characterizes the initial step of the filter recursion, and its derivation within Lemma~\ref{lem1} provides the key intuitions to obtain the state predictive $p(\btheta_t \mid \by_{1:t-1})$ and filtering $p(\btheta_t \mid \by_{1:t})$ densities, for any $t\ge 2$. Lemma \ref{lem1} states that $p(\btheta_1 \mid \by_1)$ is a \textsc{sun} density. In the following, consistent with the notation of Sect.~\ref{sec.2}, whenever $\bOmega$ is a $q \times q$ covariance matrix, the associated matrices $\bomega$ and $\bar{\bOmega}$ are defined as $\bomega=(\bOmega \odot {\bf I}_q)^{1/2}$ and $\bar{\bOmega}=\bomega^{-1}\bOmega\bomega^{-1}$, respectively. All the proofs can be found in Appendix A, and leverage conjugacy properties of the  \textsc{sun} in probit models. The first result on this property has been derived by   \citet{Durante2018} for static univariate Bayesian probit regression. Here, we take a substantially different perspective by focusing on online inference in both multivariate and time-varying probit models that require novel and non-straightforward extensions. As seen in \citet{soyer2013bayesian} and \citet{chib1998}, the increased complexity of this endeavor  typically motivates a separate treatment relative to the static univariate case. 

\vspace{-2pt}

\begin{Lemma}
	Under the dynamic probit model in  \eqref{eq1}--\eqref{eq2}, the first-step filtering distribution is
	\vspace{-4pt}
	\begin{equation}
	(\btheta_1 \mid \by_1) \sim \mbox{\textsc{sun}}_{p,m}(\bxi_{1\mid1}, \bOmega_{1\mid1}, \bDelta_{1\mid1}, \bgamma_{1\mid1}, \bGamma_{1\mid1}),
	\label{eq8}
	\end{equation}
	with parameters defined by the recursive equations
	\begin{align*}
	\bxi_{1\mid1}&=\bG_1 \ba_0, \quad \bOmega_{1\mid1}=\bG_1\bP_0\bG_1^{\intercal}+\bW_1, \\
\bDelta_{1\mid1}&=\bar{\bOmega}_{1\mid1}{\bomega}_{1\mid1}\bF_1^{\intercal} \bB_1\bs_1^{-1}, \quad \bgamma_{1\mid1}=\bs_1^{-1}\bB_1\bF_1\bxi_{1\mid1}, \\ 
\bGamma_{1\mid1}&= \bs^{-1}_1\bB_1(\bF_1\bOmega_{1\mid1} \bF^{\intercal}_1+\bV_1)\bB_1\bs^{-1}_1,
	\end{align*}
where $\bs_1= [(\bF_1\bOmega_{1\mid1} \bF^{\intercal}_1+\bV_1) \odot {\bf I}_m]^{1/2}$.
	\label{lem1}
\end{Lemma}
Hence $p(\btheta_1 \mid \by_1)$ is a \textsc{sun} density with parameters that can be obtained via tractable arithmetic expressions applied to the quantities defining  model \eqref{eq1}--\eqref{eq2}. Exploiting the results in Lemma  \ref{lem1}, the general filter updates for the multivariate dynamic probit model  can be obtained by induction for $t\geq 2$ and are presented in Theorem   \ref{teo1}.

\begin{Theorem}
	Let $(\btheta_{t-1} {\mid} \by_{1:t-1}) \sim \mbox{\textsc{sun}}_{p,m(t-1)}(\bxi_{t-1\mid t-1}, \\	
	\bOmega_{t-1\mid t-1}, \bDelta_{t-1\mid t-1}, \bgamma_{t-1\mid t-1}, \bGamma_{t-1\mid t-1})$ denote the filtering distribution at time $t-1$ under model \eqref{eq1}--\eqref{eq2}. Then, the one-step-ahead state predictive distribution at $t$ is
	\vspace{-4pt}
	\begin{align}
	&(\btheta_t \mid \by_{1:t-1}) 	\label{eq9}\\
	&{\sim} \ \mbox{\textsc{sun}}_{p,m(t{-}1)}(\bxi_{t\mid t-1}, \bOmega_{t\mid t-1}, \bDelta_{t\mid t-1}, \bgamma_{t\mid t-1}, \bGamma_{t\mid t-1}), \nonumber
	\end{align}
	        with parameters defined by the recursive equations
	        \vspace{-4pt}
	\begin{align*}
	\bxi_{t\mid t-1}&=\bG_t \bxi_{t-1\mid t-1}, \   \bOmega_{t\mid t-1}=\bG_t\bOmega_{t-1\mid t-1}\bG_t^{\intercal}+\bW_t, \\
	\bDelta_{t\mid t-1}&={\bomega}^{-1}_{t\mid t-1}\bG_t {\bomega}_{t-1\mid t-1}\bDelta_{t-1\mid t-1},\\
	\bgamma_{t\mid t-1}&=\bgamma_{t-1\mid t-1}, \quad   \ \bGamma_{t\mid t-1}=\bGamma_{t-1\mid t-1}.
	\end{align*}
Moreover, the filtering distribution at time $t$  is 
	\begin{equation}
	(\btheta_t \mid \by_{1:t}) \sim \mbox{\textsc{sun}}_{p,m t}(\bxi_{t\mid t}, \bOmega_{t\mid t}, \bDelta_{t\mid t}, \bgamma_{t\mid t}, \bGamma_{t\mid t}),
	\label{eq10}
	\end{equation}
			with parameters defined by the recursive equations
	\begin{align*}
\bxi_{t\mid t}&= \bxi_{t\mid t-1}, \quad 	\bOmega_{t\mid t}= \bOmega_{t\mid t-1},\\
\bDelta_{t\mid t}&= [\bDelta_{t\mid t-1} , \bar{\bOmega}_{t\mid t}{\bomega}_{t\mid t}\bF_t^{\intercal}\bB_t\bs^{-1}_{t}], \\
\bgamma_{t\mid t}&= [\bgamma_{t\mid t-1}^{\intercal}, \bxi_{t\mid t}^{\intercal}\bF_t^{\intercal}\bB_t\bs^{-1}_{t}]^{\intercal},
	\end{align*}
 and $\bGamma_{t\mid t}$ is a full-rank correlation matrix having blocks $\bGamma_{t\mid t [11]}=\bGamma_{t\mid t-1}$, $\bGamma_{t\mid t [22]}=\bs^{-1}_t\bB_t(\bF_t\bOmega_{t\mid t} \bF^{\intercal}_t{+}\bV_t)\bB_t\bs^{-1}_t$ and $\bGamma_{t\mid t [21]}=\bGamma_{t\mid t [12]}^{\intercal}=\bs_t^{-1}\bB_t \bF_t \bomega_{t\mid t}\bDelta_{t\mid t-1}$, where $\bs_t$ is defined as $\bs_t=[(\bF_t\bOmega_{t\mid t} \bF^{\intercal}_t+\bV_t) \odot {\bf I}_m]^{1/2}$.
	\label{teo1}
\end{Theorem}

As shown in Theorem \ref{teo1}, online prediction and filtering in the multivariate dynamic probit model  \eqref{eq1}--\eqref{eq2} proceeds by iterating between equations \eqref{eq9} and \eqref{eq10} as new observations stream in with time $t$. Both steps are based on closed-form distributions and rely  on analytical expressions for recursive updating of the corresponding parameters as a function of the previous ones, thus providing an analog of the classical Kalman filter.

We also provide closed-form expressions for the predictive density of the multivariate binary response data $\by_{t}$. Indeed, the prediction of $\by_{t} \in \{0;1\}^{m}$ given the data $\by_{1:t-1}$, is a primary goal in applications of dynamic probit models. In our setting, this task requires the derivation of the predictive density $p(\by_{t} \mid \by_{1:t-1})$ which coincides, under  \eqref{eq1}--\eqref{eq2}, with $\int \Phi_m(\bB_t\bF_t\btheta_{t}; \bB_t\bV_t\bB_t)p(\btheta_t \mid \by_{1:t-1}) \mbox{d} \btheta_t$, where $p(\btheta_t \mid \by_{1:t-1})$ is the state predictive density from  \eqref{eq9}. Corollary  \ref{cor1} shows that $p(\by_{t} \mid \by_{1:t-1})$ has an explicit form.

\begin{Corollary}
	Under model \eqref{eq1}--\eqref{eq2}, the observation predictive density  $p(\by_{t} \mid \by_{1:t-1})$ is
	\begin{equation}
	p(\by_{t} \mid \by_{1:t-1})=\frac{\Phi_{m  t}(\bgamma_{t\mid t}; \bGamma_{t\mid t})}{\Phi_{m (t-1)}(\bgamma_{t\mid t-1}; \bGamma_{t\mid t-1})},
	\label{eq11}
	\end{equation}
	for every time $t$, with parameters $\bgamma_{t\mid t}$, $\bGamma_{t\mid t}$, $\bgamma_{t\mid t-1}$ and $\bGamma_{t\mid t-1}$, defined as in Theorem \ref{teo1}.
	\label{cor1}
\end{Corollary}
Hence, the evaluation of probabilities of future events is possible via explicit calculations after marginalizing out analytically the states with respect to their predictive density. As is clear from \eqref{eq11}, this  requires the calculation of Gaussian cumulative distribution functions whose dimension increases with $t$ and $m$. Efficient evaluation of such integrals is possible for small-to-moderate $t$ and $m$ via recent methods \citep{botev_2017}, but  this solution is impractical for large  $t$ and $m$, as seen in Table~\ref{table1}. In Sect.~\ref{sec.4}, we develop novel Monte Carlo strategies to address this issue and enhance scalability. This is done by exploiting  Theorem \ref{teo1} to improve current solutions.

\vspace{-9pt}

\subsection{Smoothing distribution}
\label{subsec:Smoothing}
\vspace{-6pt}
We now consider smoothing distributions. In this case,  the focus is on the distribution of the entire states' sequence $\btheta_{1:n}$, or  a subset of it, given all data $\by_{1:n}$. Theorem \ref{thm:JointSmoothing} shows that also the smoothing density $p(\btheta_{1:n} \mid \by_{1:n})$ belongs to the \textsc{sun} family. Direct consequences of this result, involving marginal smoothing and marginal likelihoods are reported in Corollaries \ref{cor:MarginalSmoothing} and \ref{cor:MarginalLikelihood}.

Before stating the result, let us first  introduce the two block-diagonal matrices, $\bD$ and $\bLambda$,  with dimensions $(m n) \times (p  n)$ and $(m  n) \times (m n)$ respectively, and diagonal blocks $\bD_{[ss]}=\bB_s\bF_s \in \mathbb{R}^{m \times p}$ and $\bLambda_{[ss]}=\bB_s\bV_s\bB_s \in \mathbb{R}^{m \times m}$, for every time point $s=1, \ldots, n$. Moreover, let $\bxi$ and $\bOmega$ denote the mean and covariance matrix of the multivariate Gaussian distribution for $\btheta_{1:n}$ induced by the state equations. Under  \eqref{eq2}, $\bxi$ is a $p n \times 1$ column vector obtained by stacking the $p$-dimensional blocks $\bxi_{[s]}=\mathbb{E}(\btheta_s)=\bG^{s}_1 \ba_0 \in \mathbb{R}^p$ for every $s=1, \ldots, n$, with $\bG_{1}^{s}=\bG_s \cdots \bG_1$. Similarly, letting $\bG_{l}^{s}=\bG_s \cdots \bG_l$, also the $(p  n) \times (p  n)$ covariance matrix  $\bOmega$ has a block structure with $(p \times p)$-dimensional blocks $\bOmega_{[ss]}=\mbox{var}(\btheta_s)=\bG^{s}_1 \bP_0 \bG^{s\intercal}_1+\sum_{l=2}^s\bG^{s}_l \bW_{l-1}\bG^{s\intercal}_l+\bW_s$, for $s=1, \ldots,n$, and  $\bOmega_{[sl]}=\bOmega^{\intercal}_{[ls]}=\mbox{cov}(\btheta_s, \btheta_l)=\bG_{l+1}^{s}\bOmega_{[ll]}$, for $s>l$.

\begin{Theorem}
	\label{thm:JointSmoothing}
	Under model \eqref{eq1}--\eqref{eq2}, the joint smoothing distribution is
	\vspace{-5pt}
	\begin{align}
		&(\btheta_{1:n} \mid \by_{1:n}) \label{eq:JointSmoothing} \\	
	&\sim \mbox{\textsc{sun}}_{p n, m  n}(\bxi_{1:n\mid n},\bOmega_{1:n\mid n}, \bDelta_{1:n\mid n},  \bgamma_{1:n\mid n}, \bGamma_{1:n\mid n}), \nonumber
	\end{align}
with parameters defined as
	\vspace{-5pt}
\begin{align*}
\bxi_{1:n\mid n}&=\bxi, \quad \bOmega_{1:n\mid n}=\bOmega,  \quad \bDelta_{1:n\mid n}=\bar{\bOmega}\bomega\bD^{\intercal}\bs^{-1},\\	
\bgamma_{1:n\mid n}&=\bs^{-1}\bD\bxi, \quad  \bGamma_{1:n\mid n}=\bs^{-1}(\bD\bOmega\bD^{\intercal}+\bLambda)\bs^{-1},
\end{align*}
where $\bs=[(\bD\bOmega\bD^{\intercal}+\bLambda) \odot \mbox{\bf I}_{m n}]^{1/2}$.
\end{Theorem}

Since the \textsc{sun} is closed under marginalization  and linear combinations, it follows from Theorem \ref{thm:JointSmoothing} that the smoothing distribution for any combination of states  is still a \textsc{sun}. In particular, direct application of the results in \citet[][Sect.~7.1.2]{azzalini_2013} yields the  marginal smoothing distribution for any state $\btheta_t$ reported in Corollary \ref{cor:MarginalSmoothing}.

\begin{Corollary}
	\label{cor:MarginalSmoothing}
	Under the model in  \eqref{eq1}--\eqref{eq2}, the marginal smoothing distribution at any time $t \leq n$ is
	\begin{equation}
	(\btheta_{t} \mid \by_{1:n}) \sim \mbox{\textsc{sun}}_{p, m n}(\bxi_{t \mid n},\bOmega_{t \mid n}, \bDelta_{t \mid n}, \bgamma_{t \mid n}, \bGamma_{t \mid n}),
		\label{eq:MarginalSmoothing}
	\end{equation}
	with parameters defined as
\begin{align*}
\bxi_{t \mid n}&=\bxi_{[t]}, \quad \ \ \	\bOmega_{t \mid n}=\bOmega_{[tt]}, \quad \bDelta_{t \mid n}=\bDelta_{1:n\mid n[t]}, \\
 \bgamma_{t \mid n}&=\bgamma_{1:n\mid n}, \quad  \bGamma_{t \mid n}=\bGamma_{1:n\mid n},
\end{align*}	
where $\bDelta_{1:n\mid n[t]}$ defines the $t$-th block of $p$ rows in $\bDelta_{1:n\mid n}$. When $t=n$, (\ref{eq:MarginalSmoothing}) gives the filtering distribution at  $n$.
\end{Corollary}

Another important consequence of Theorem \ref{thm:JointSmoothing} is the availability of a closed-form expression for the marginal likelihood $p(\by_{1:n})$, which is provided in Corollary \ref{cor:MarginalLikelihood}.
\begin{Corollary}
	\label{cor:MarginalLikelihood}
	Under model \eqref{eq1}--\eqref{eq2}, the marginal likelihood is
	$p(\by_{1:n})=\Phi_{m n}(\bgamma_{1:n\mid n};\bGamma_{1:n\mid n}),$
	with $\bgamma_{1:n\mid n}$ and $\bGamma_{1:n\mid n}$ defined as in Theorem \ref{thm:JointSmoothing}.
\end{Corollary}

This closed-form result is useful in several contexts, including estimation of unknown system parameters via marginal likelihood maximization, and full Bayesian inference through \textsc{mcmc} or variational inference methods.

\section{Inference via Monte Carlo methods}
\label{sec.4}
As discussed in Sects.~\ref{sec.2} and \ref{sec.3}, inference without sampling from \eqref{eq9}, \eqref{eq10} or \eqref{eq:JointSmoothing} is, theoretically, possible. Indeed, since the \textsc{sun} densities of the filtering, predictive and smoothing distributions can be obtained from Theorems \ref{teo1}--\ref{thm:JointSmoothing}, the main functionals of interest can be  computed via closed-form expressions  \citep{arellano_2006,azzalini_2010, gupta_2013,azzalini_2013} or by relying on numerical integration. However, these strategies require evaluations  of multivariate Gaussian cumulative distribution functions, which tend to be impractical as $t$ grows or when the focus is on complex functionals.

In such situations, Monte Carlo integration provides an accurate solution to evaluate the generic functionals $\mathbb{E}[g(\btheta_{t}) \mid \by_{1:t}]$, $\mathbb{E}[g(\btheta_{t}) \mid \by_{1:t-1}]$  and $\mathbb{E}[g(\btheta_{1:n}) \mid \by_{1:n}]$ for the filtering, predictive and smoothing distribution via 
\begin{equation*}
\frac{1}{R}\sum_{r=1}^Rg(\btheta^{(r)}_{t\mid t}), \quad \frac{1}{R}\sum_{r=1}^Rg(\btheta^{(r)}_{t\mid t-1}),  \quad \frac{1}{R}\sum_{r=1}^Rg(\btheta^{(r)}_{1:n\mid n}),
\end{equation*}
with  $\btheta^{(r)}_{t\mid t}$,  $\btheta^{(r)}_{t\mid t-1}$ and $\btheta^{(r)}_{1:n\mid n}$ sampled from $p(\btheta_{t} \mid \by_{1:t})$, $p(\btheta_{t} \mid \by_{1:t-1})$ and $p(\btheta_{1:n} \mid \by_{1:n})$, respectively. For example, if the evaluation of \eqref{eq11} is  demanding, the observations predictive density  can be easily computed as $\sum_{r=1}^R \Phi_m(\bB_{t}\bF_{t}\btheta^{(r)}_{t \mid t-1}; \bB_{t}\bV_{t}\bB_{t})/R$.

\begin{algorithm*}[t!]
	\caption{\normalsize \fontfamily{lmss}\selectfont  Independent and identically distributed sampling from $p(\btheta_{1:n} \mid \by_{1:n})$} 
	\label{algo1}
	\vspace{5pt}
\fontfamily{lmss}\selectfont 
	\begin{description}
		\item{{\bf [1]}  Sample $\bU^{(1)}_{0 \ 1:n\mid n}, \ldots, \bU^{(R)}_{0 \ 1:n\mid n}$ independently from  a $\mbox{N}_{p n}({\bf 0},\bar{\bOmega}_{1:n\mid n}- \bDelta_{1:n\mid n}\bGamma_{1:n\mid n}^{-1}\bDelta_{1:n\mid n}^{\intercal})$.}
		\item{{\bf [2]} Sample $\bU^{(1)}_{1 \ 1:n\mid n}, \ldots, \bU^{(R)}_{1 \ 1:n\mid n}$ independently from  a $\textsc{tn}_{m  n}({\bf 0},\bGamma_{1:n\mid n}; \mathbb{A}_{\bgamma_{1:n\mid n}})$.}
		\item{{\bf [3]} Compute $\btheta^{(1)}_{1:n\mid n}, \ldots, \btheta^{(R)}_{1:n\mid n}$ via $\btheta^{(r)}_{1:n\mid n}=\bxi_{1:n\mid n}+\bomega_{1:n\mid n}(\bU^{(r)}_{0 \ 1:n\mid n}+\bDelta_{1:n\mid n} \bGamma_{1:n\mid n}^{-1} \bU^{(r)}_{1 \ 1:n\mid n})$, for $r=1, \ldots, R$.}
	\end{description}
\end{algorithm*}

To be implemented, the above approach requires an efficient strategy to sample  from \eqref{eq9}, \eqref{eq10} and \eqref{eq:JointSmoothing}. Exploiting the \textsc{sun} properties and  recent results in \citet{botev_2017}, an algorithm to draw independent and identically distributed samples from the exact  \textsc{sun} distributions in \eqref{eq9}, \eqref{eq10} and \eqref{eq:JointSmoothing} is developed in Sect.~\ref{sec.41}. As illustrated in Sect.~\ref{sec.5}, such a technique is  more accurate than state-of-the-art methods and can be efficiently implemented in small-to-moderate dimensional time series. In Sect.~\ref{sec.42} we develop, instead, novel sequential Monte Carlo schemes that allow scalable online learning in high dimensional settings and have optimality properties \citep{doucet2000}  which shed new light also on  existing strategies  \citep[e.g,][]{andrieu2002}.

\subsection{Independent identically distributed sampling}
\label{sec.41}
As discussed in Sect.~\ref{sec_1}, \textsc{mcmc} and sequential Monte Carlo methods to sample from $p(\btheta_{t} \mid \by_{1:t})$, $p(\btheta_{t} \mid \by_{1:t-1})$ and $p(\btheta_{1:n} \mid \by_{1:n})$ are available. However, the commonly recommended practice, if feasible, is to rely on independent and identically distributed (i.i.d.) samples. Here, we derive a Monte Carlo algorithm to address this goal with a main focus on the smoothing distribution, and discuss direct modifications to allow sampling also in the filtering and predictive case. Indeed, Monte Carlo inference is particularly suitable for batch settings, although, as discussed later,  the proposed routine is practically useful also when the focus is on filtering and predictive distributions, since i.i.d.\ samples are simulated rapidly, for each $t$, in small-to-moderate dimensions.

Exploiting the closed-form expression of the smoothing distribution in Theorem \ref{thm:JointSmoothing}, and the additive representation \eqref{eq7} of the \textsc{sun}, i.i.d. samples for $\btheta_{1:n\mid n}$ from the smoothing distribution  \eqref{eq:JointSmoothing} can be  obtained via a linear combination between independent samples from $(p n)$-variate Gaussians and $(m n)$-variate truncated normals. Algorithm \ref{algo1} provides the detailed pseudo-code for this novel strategy, whose outputs are i.i.d. samples from the joint smoothing density  $p(\btheta_{1:n} \mid \by_{1:n})$. Here, the most computationally intensive step is the sampling from $\textsc{tn}_{m  n}({\bf 0},\bGamma_{1:n\mid n}; \mathbb{A}_{\bgamma_{1:n\mid n}})$, which denotes the multivariate normal distribution $\mbox{N}_{m n}({\bf 0},\bGamma_{1:n\mid n})$ truncated to the region $\mathbb{A}_{\bgamma_{1:n\mid n}}=\{\bu_1 \in \mathbb{R}^{m n}: \bu_1+\bgamma_{1:n\mid n}>0\}$. In fact, although efficient Hamiltonian Monte Carlo solutions are available \citep{pakman_2014}, these strategies do not provide independent samples. More recently, an accept-reject method based on minimax tilting has been proposed by  \citet{botev_2017} to improve the acceptance rate of classical rejection sampling, while avoiding mixing issues of \textsc{mcmc}. This routine is available in the \texttt{R} library \texttt{TruncatedNormal} and allows efficient sampling from multivariate truncated normals with a dimension of few hundreds, thereby providing effective Monte Carlo inference via Algorithm~\ref{algo1}  in small-to-moderate dimensional time series where $m n$ is of the order of few hundreds.

Clearly, the availability of an i.i.d.\ sampling scheme from the smoothing distribution overcomes the need of \textsc{mcmc} methods and particle smoothers. The first set of strategies usually faces mixing or time-inefficiency issues, especially in imbalanced binary settings \citep{Johndrow2018}, whereas the second class of routines tends to be computationally intensive and subject to   particles degeneracy  \citep{Doucet2009}.

When the focus is on Monte Carlo inference for the marginal smoothing density $p(\btheta_t \mid \by_{1:n})$  at a specific time $t$, Algorithm \ref{algo1} requires minor adaptations relying again on the additive representation of the \textsc{sun} in  \eqref{eq:MarginalSmoothing}, under similar arguments considered for the joint smoothing setting. This latter routine can be also used to sample from the filtering distribution in \eqref{eq10} by applying such a scheme with  $n=t$  to obtain i.i.d.\ samples for $\btheta_{t\mid t}$ from $p(\btheta_{t} \mid \by_{1:t})$. Leveraging  realizations from the filtering distribution at time $t-1$, i.i.d.\ samples for $\btheta_{t \mid t-1}$ from the  predictive density $p(\btheta_{t} \mid \by_{1:t-1})$, can be simply obtained via the  direct application of  \eqref{eq2} which provides samples for  $\btheta_{t\mid t-1}$ from  $\mbox{N}_p(\bG_{t}\btheta_{t-1\mid t-1}, \bW_{t})$. As a result, efficient Monte Carlo inference in small-to-moderate dimensional dynamic probit models is possible also for  filtering and predictive distributions.

\subsection{Sequential Monte Carlo sampling}
\label{sec.42}
When the dimension of the dynamic probit model  \eqref{eq1}--\eqref{eq2} grows, sampling from multivariate truncated Gaussians in Algorithm \ref{algo1} might yield computational bottlenecks \citep{botev_2017}. This is particularly likely to occur in series monitored on a fine time grid. Indeed, in several applications, the number of time  series $m$ is typically small, whereas the length of the time window can be large. To address this issue and allow scalable online filtering and prediction  also in large $t$ settings, we first derive in Sect.~\ref{sec.421} a particle filter which exploits the \textsc{sun} results to obtain optimality properties, in the sense of \citet{doucet2000}. Despite covering a gap in the literature on dynamic probit models, as clarified in Sects.~\ref{sec.421} and \ref{sec.422},  such a strategy is amenable to further improvements since it induces unnecessary autocorrelation in the Gaussian part of the \textsc{sun} generative representation. Motivated by this consideration and by the additive structure of the \textsc{sun} filtering distribution, we further develop in Sect.~\ref{sec.422} a partially collapsed sequential Monte Carlo procedure which recursively samples via lookahead methods  \citep{lin2013} only the multivariate truncated normal term in the  \textsc{sun} additive representation, while keeping the Gaussian component exact. As outlined in Sect.~\ref{sec.422}, such a broad class of partially collapsed lookahead particle filters comprises, as a special case, the Rao--Blackwellized particle filter developed by   \citet{andrieu2002}. This provides novel theoretical support to the notable performance of such a strategy, which was originally motivated, in the context of dynamic probit models, also by the lack of a closed-form optimal particle filter for the states.

\begin{algorithm*}[t]
	\caption{\normalsize \fontfamily{lmss}\selectfont  ``Optimal'' particle filter to sample from $p(\btheta_t \mid \by_{1:t})$, for $t=1, \ldots, n$ [\textsc{auf} version]} 
	\label{algo2}
	\vspace{5pt}
	\fontfamily{lmss}\selectfont 
	\For(){$t$ \mbox{from} $1$ to $n$}
	{
	{\bf [1]} Compute the weights $w_t^{(r)}=p(\by_t \mid {\btheta}_{t-1}={\btheta}_{t-1 \mid t-1}^{(r)})$ for $r=1, \ldots, R$, by applying equation \eqref{eq14}.	\\
	{\bf [2]} Resample updated particles $\bar{\btheta}^{(1)}_{t-1\mid t-1}, \ldots, \bar{\btheta}^{(R)}_{t-1\mid t-1}$ from $\sum_{r=1}^{R} w_t^{(r)}\delta_{{\btheta}_{t-1 \mid t-1}^{(r)}}$.	\\
	\vspace{-1pt}
	\For(){$r$ \mbox{from} $1$ to $R$}{
				\vspace{1pt}
			   {\bf [3]}  Set $\bxi^{(r)}_{t \mid t,t-1}=\bG_t \bar{\btheta}_{t-1\mid t-1}^{(r)}$ and $\bgamma_{t\mid t,t-1}^{(r)}= \bc_t^{-1}\bB_t\bF_t \bxi^{(r)}_{t \mid t,t-1}$. Then, simulate ${\btheta}^{(r)}_{t\mid t}$ from \eqref{eq13}, as follows:
			\vspace{-3pt}
			\begin{description}
				\item{{\bf [3.1]} Sample $\bU^{(r)}_{0 \ t\mid t}$ from  a $\mbox{N}_p({\bf 0},\bar{\bOmega}_{t\mid t, t-1}- \bDelta_{t\mid t, t-1}\bGamma_{t\mid t, t-1}^{-1}\bDelta_{t\mid t, t-1}^{\intercal})$.}
				\vspace{2pt}
				 \item{{\bf  [3.2]} Sample $\bU^{(r)}_{1 \ t\mid t}$  from  a $\textsc{tn}_{m}({\bf 0},\bGamma_{t\mid t, t-1}; \mathbb{A}_{\bgamma_{t\mid t,t-1}^{(r)}})$.}
				 \item{{\bf  [3.3]} Compute ${\btheta}^{(r)}_{t\mid t}=\bxi^{(r)}_{t \mid t,t-1}+\bomega_{t\mid t, t-1}(\bU^{(r)}_{0 \ t\mid t}+\bDelta_{t\mid t, t-1} \bGamma_{t\mid t,t-1}^{-1} \bU^{(r)}_{1 \ t\mid t})$.}
			\end{description}
			\vspace{-5pt}
	}
\vspace{-3pt}	}
\end{algorithm*}

\vspace{8pt}
\subsubsection{``Optimal'' particle filter}
\label{sec.421}
The first proposed strategy belongs to the class of sequential importance sampling-resampling (\textsc{sisr}) algorithms that provide default strategies in particle filtering \citep[e.g.,][]{doucet2000, doucet2001,durbin2012}. For each time $t$, these routines sample $R$ trajectories for $\btheta_{1:t|t}$ from $p(\btheta_{1:t} \mid \by_{1:t})$, known as \emph{particles}, conditioned on those produced at $t-1$, by iterating, in time, between the two  steps summarized below.

\begin{description}[leftmargin=0cm]
\item{\noindent{\bf 1. Sampling}. Let $\btheta_{1:t-1\mid t-1}^{(1)},\ldots,\btheta_{1:t-1\mid t-1}^{(R)}$  be the trajectories of the  particles  at time $t-1$, and denote with $\pi(\btheta_{t} \mid  \btheta_{1:t-1}, \by_{1:t})$ the proposal. Then, for $r=1, \ldots, R$\\
\begin{description}
			\item{[1.a] Sample $\bar{\btheta}_{t \mid t}^{(r)}$ from $\pi({\btheta}_{t} \mid \btheta_{1:t-1 \mid t-1}^{(r)}, \by_{1:t} )$ and set 
\begin{equation*}			
\bar{\btheta}_{1:t \mid t}^{(r)} = (\btheta_{1:t-1 \mid t-1}^{(r)\intercal}, \bar{\btheta}_{t \mid t}^{(r) \intercal} )^{\intercal}.
\end{equation*}	}
			\item{[1.b] Set $w^{(r)}_t= w_t(\bar{\btheta}_{1:t \mid t}^{(r)})$, with
\begin{equation*}			
w_t(\bar{\btheta}_{1:t \mid t}^{(r)})  \propto \frac{p(\by_t \mid \bar{\btheta}_{t \mid t}^{(r)})p(\bar{\btheta}_{t \mid t}^{(r)} \mid \btheta_{t-1 \mid t-1}^{(r)})}{\pi( \bar{\btheta}^{(r)}_{t \mid t} \mid \btheta_{1:t-1 \mid t-1}^{(r)}, \by_{1:t} )},
\end{equation*}			
and normalize the weights, so that their sum is $1$.\\}
\end{description}}		

\item{\noindent{\bf 2. Resampling}. For $r=1, \ldots, R$, sample updated particles' trajectories $\btheta_{1:t\mid t}^{(1)},\ldots,\btheta_{1:t\mid t}^{(R)}$ from $\sum_{r=1}^{R} w_t^{(r)}\delta_{\bar{\btheta}_{1:t \mid t}^{(r)}}$.}
\end{description}

From these particles, functionals of the filtering density $p(\btheta_t \mid \by_{1:t})$ can be computed  using the terminal values $\btheta_{t|t}$ of each particles' trajectory for $\btheta_{1:t|t}$. Note that in point [1.a] we have presented the general formulation of \textsc{sisr}, where the importance density  $\pi(\btheta_{t} \mid  \btheta_{1:t-1}, \by_{1:t})$ can, in principle, depend on the whole trajectory $\btheta_{1:t-1}$ \citep[Sect. 12.3]{durbin2012}.

As is clear from the above steps, the performance of \textsc{sisr} relies on the choice of $\pi(\btheta_{t} \mid  \btheta_{1:t-1}, \by_{1:t})$. Such a density should allow tractable sampling along with efficient evaluation of the importance weights, and should be also  carefully specified to propose effective candidate samples. Recalling \citet{doucet2000}, the optimal proposal is $\pi(\btheta_{t} \mid  \btheta_{1:t-1}, \by_{1:t})=p( \btheta_{t} \mid \btheta_{t-1}, \by_{t} )$, with importance weights $w_t \propto p(\by_t \mid \btheta_{t-1} )$. Indeed, conditioned on $\btheta_{1:t-1|t-1}$ and $\by_{1:t}$, this choice minimizes the variance of the weights, thus limiting  degeneracy issues and improving mixing. Unfortunately, in several dynamic models, tractable sampling from $p( \btheta_{t} \mid \btheta_{t-1}, \by_{t} )$ and the direct evaluation of $p(\by_t \mid \btheta_{t-1} )$ is not possible  \citep{doucet2000}. As outlined in Corollary \ref{prop1}, this is not the case for dynamic probit models. In particular, by leveraging the proof of Theorem \ref{teo1} and the closure properties of the \textsc{sun}, sampling from $p( \btheta_{t} \mid \btheta_{t-1}, \by_{t} )$ is straightforward and $p(\by_t \mid \btheta_{t-1} )$ has a simple  form. 

\begin{Corollary}
	For every time $t=1,\ldots,n$, the optimal importance distribution under model \eqref{eq1}--\eqref{eq2}  is
	\vspace{-2pt}
		\begin{align}
	&( \btheta_{t} \mid \btheta_{t-1}, \by_{t} ) \label{eq13}\\
	& \sim\mbox{\textsc{sun}}_{p,m}(\bxi_{t \mid t,t-1}, \bOmega_{t\mid t,t-1}, \bDelta_{t\mid t, t-1},\bgamma_{t\mid t, t-1}, \bGamma_{t\mid t, t-1}),  \nonumber 	
		\end{align}
		whereas the importance weights are
		\vspace{-2pt}
	\begin{equation}
p(\by_{t} \mid \btheta_{t-1}) = \Phi_m(\bgamma_{t\mid t, t-1}; \bGamma_{t\mid t, t-1}),\label{eq14}
	\end{equation}
	with parameters defined by the recursive equations
	\begin{align*}
	\bxi_{t \mid t,t-1}&=\bG_t \btheta_{t-1}, \quad \bOmega_{t\mid t,t-1}=\bW_t,\\
	\bDelta_{t\mid t, t-1}&=\bar{\bOmega}_{t\mid t,t-1}\bomega_{t\mid t,t-1}\bF_t^{\intercal}\bB_t\bc_t^{-1},\\
	\bgamma_{t\mid t, t-1}&=\bc_t^{-1}\bB_t\bF_t \bxi_{t \mid t,t-1}, \\
	\bGamma_{t\mid t, t-1}&=\bc_t^{-1}\bB_t\left(\bF_t \bOmega_{t\mid t,t-1}\bF_t^{\intercal}{+}\bV_t\right) \bB_t \bc_t^{-1},
	\end{align*}
	where $\bc_t = \left[(\bF_t \bOmega_{t\mid t,t-1}\bF_t^{\intercal}+\bV_t)\odot\bI_m\right]^{1/2}$.
	\label{prop1}
\end{Corollary}

As clarified in Corollary \ref{prop1}, the weights $p(\by_{t} \mid \btheta_{t-1})$ for the generated trajectories  are available analytically in (\ref{eq14}) and do not depend on the sampled values of the particle at time $t$. This allows the implementation of the more efficient auxiliary particle filter (\textsc{auf}) \citep{pitt1999} by simply reversing the order of the sampling and resampling steps, thereby obtaining a performance gain \citep{andrieu2002}. Algorithm \ref{algo2} illustrates the pseudo-code of the proposed ``optimal'' auxiliary  filter, which exploits the additive representation of the \textsc{sun} and Corollary \ref{prop1}. Note that, unlike for Algorithm \ref{algo1}, such a sequential sampling strategy requires to sample at each step from a truncated normal whose dimension  does not depend on $t$, thus facilitating scalable sequential inference in large $t$ studies. Samples from the predictive distribution can be obtained from those of the filtering as discussed  in Sect.~\ref{sec.41}.

Despite having optimality properties, a close inspection of Algorithm \ref{algo2} shows that the states' particles  at $t-1$ affect both the Gaussian component, via $\bxi_{t \mid t,t-1}$, and the truncated normal term, via $\bgamma_{t \mid t,t-1}$, in the \textsc{sun} additive representation of $(\btheta_t \mid \by_{1:t})$. Although the autocorrelation in the multivariate truncated normal samples is justified by the computational intractability of this variable in high dimensions, inducing serial dependence also in the Gaussian terms seems  unnecessary, as these quantities are tractable and their dimension does not depend on $t$; see Theorem \ref{teo1}. This suggests that a strategy which sequentially updates only the truncated normal term, while maintaining the Gaussian part exact, could further improve the performance of Algorithm \ref{algo2}. This new particle filter is derived in Sect.~\ref{sec.422}, inheriting  also lookahead ideas  \citep{lin2013}.

\subsubsection{Partially collapsed lookahead particle filter}
\label{sec.422}
\vspace{-5pt}
As anticipated within Sect.~\ref{sec.42}, the most computationally intensive step to draw i.i.d.\ samples from the filtering distribution is sampling from the multivariate truncated normal $\bU_{1 \ 1:t\mid t}\sim \textsc{tn}_{m t}(\boldsymbol{0},\bGamma_{1:t\mid t}; \mathbb{A}_{\bgamma_{1:t\mid t}})$ in Algorithm \ref{algo1}. Here, we present a  class of procedures to sequentially generate these samples, which are then combined with realizations from the exact Gaussian component in the \textsc{sun} additive representation, thus producing samples from the filtering distribution. With this goal in mind, define the region $\mathbb{A}_{\by_{s:t}}= \{\bz \in \mathbb{R}^{m(t-s+1)}: (2\by_{s:t}-\boldsymbol{1})\odot \bz>\boldsymbol{0}\}$ for every  $s=1, \ldots, t$, and let $\bV_{1:t}$ be the $(mt) \times (mt)$ block-diagonal matrix having blocks $\bV_{[ss]}=\bV_s$, for $s=1, \ldots, t$. Moreover, denote with $\bB_{s:t}$ and $\bF_{s:t}$ two block-diagonal matrices of dimension $[m(t-s+1)]\times[m (t-s+1)]$ and $[m (t-s+1)]\times[p (t-s+1)]$, respectively, and diagonal blocks $\bB_{s:t [ll]} = \bB_{s+l-1}$ and $\bF_{s:t [ll]}=\bF_{s+l-1}$ for $l=1,\ldots,t-s+1$. Exploiting this notation and adapting results in Sect.~\ref{subsec:Smoothing} to the case  $n=t$, it  follows from standard properties of multivariate truncated normals \citep{horrace2005} that
\vspace{-2pt}
\begin{equation}
	\bU_{1 \ 1:t\mid t} \stackrel{\mbox{\scriptsize d}}= -\bgamma_{1:t\mid t} + \bs_{1:t\mid t}^{-1}\bB_{1:t}{\bz}_{1:t\mid t},
	\label{eq16}
\end{equation}
\vspace{-2pt}
with ${\bz}_{1:t\mid t} \sim \textsc{tn}_{m t}(\bF_{1:t}\bxi_{1:t\mid t}, \bF_{1:t}\bOmega_{1:t\mid t}\bF_{1:t}^{\intercal}+\bV_{1:t}; \mathbb{A}_{\by_{1:t}})$ and $\bs_{1:t\mid t} = [(\bD\bOmega_{1:t\mid t}\bD^{\intercal}+\bLambda) \odot \mbox{\bf I}_{m t}]^{1/2}$, where $\bD$ and $\bLambda$ are defined as in Sect.~\ref{subsec:Smoothing}, setting $n=t$. Note that the multivariate truncated normal distribution for ${\bz}_{1:t\mid t}$ actually coincides with the conditional distribution of $\bz_{1:t}$ given $\by_{1:t}$ under model \eqref{eq3}--\eqref{eq5}. Indeed, by marginalizing out $\btheta_{1:t}$ in $p(\bz_{1:t} \mid \btheta_{1:t})=\prod_{s=1}^t\phi_m(\bz_{s}-\bF_s\btheta_{s};\bV_s)=\phi_{m t}(\bz_{1:t}-\bF_{1:t}\btheta_{1:t}; \bV_{1:t})$ with respect to its multivariate normal distribution derived in the proof of Theorem~\ref{thm:JointSmoothing}, we have  $p(\bz_{1:t}) = \phi_{m t}(\bz_{1:t}-\bF_{1:t}\bxi_{1:t\mid t}; \bF_{1:t}\bOmega_{1:t\mid t}\bF_{1:t}^\intercal+\bV_{1:t})$ and, as a direct consequence, we obtain
\vspace{-2pt}
\begin{align*}
p(\bz_{1:t}\mid \by_{1:t}) &\propto p(\bz_{1:t})p(\by_{1:t} \mid \bz_{1:t}), \\
& \propto  p(\bz_{1:t})\mathbbm{1}[ (2\by_{1:t}-\boldsymbol{1})\odot \bz_{1:t}>\boldsymbol{0}],
\end{align*}
\vspace{-2pt}
which is the kernel of a $\textsc{tn}_{m t}(\bF_{1:t}\bxi_{1:t\mid t}, \bF_{1:t}\bOmega_{1:t\mid t}\bF_{1:t}^{\intercal}+\bV_{1:t}; \mathbb{A}_{\by_{1:t}})$ density.

\vspace{-1pt}
The above analytical derivations clarify that in order to sample recursively from $\bU_{1 \ 1:t\mid t}$ it is sufficient to apply equation \eqref{eq16} to  sequential realizations  of $\bz_{1:t \mid t}$ from the joint conditional density $p(\bz_{1:t}\mid \by_{1:t})$, induced by model \eqref{eq3}--\eqref{eq5}, after collapsing out   $\btheta_{1:t}$. While basic \textsc{sisr} algorithms for $p(\bz_{1:t}\mid \by_{1:t})$, combined with the exact  sampling from the Gaussian component $\bU_{0 \ t\mid t}$, are expected to yield an improved performance relative to the particle filter developed in Sect.~\ref{sec.421}, here we adapt an even broader class of lookahead particle filters \citep{lin2013} --- which includes the basic \textsc{sisr} as a special case. To introduce the general lookahead idea note that $p(\bz_{1:t}\mid \by_{1:t})=p(\bz_{t-k+1:t} \mid \bz_{1:t-k}, \by_{1:t})p(\bz_{1:t-k} \mid  \by_{1:t})$, where $k$ is a pre-specified delay offset. Moreover, as a direct consequence of the dependence structure displayed in Fig.~\ref{F_mod}, we also have that $p(\bz_{t-k+1:t} \mid \bz_{1:t-k}, \by_{1:t})=p(\bz_{t-k+1:t} \mid \bz_{1:t-k}, \by_{t-k+1:t})$ for any generic $k$. Hence, to sequentially generate realizations of $\bz_{1:t \mid t}$ from $p(\bz_{1:t}\mid \by_{1:t})$, we can first sample $\bz_{1:t-k \mid t}$ from $p(\bz_{1:t-k} \mid  \by_{1:t})$  by extending, via \textsc{sisr},  the trajectory $\bz_{1:t-k-1 \mid t-1}$ with optimal proposal $p(\bz_{t-k}\mid \bz_{1:t-k-1}=\bz_{1:t-k-1 \mid t-1}, \by_{t-k:t})$, and then draw the last $k$ terms in $\bz_{1:t \mid t}$ from $p(\bz_{t-k+1:t} \mid \bz_{1:t-k}=\bz_{1:t-k \mid t}, \by_{t-k+1:t})$. Note that when $k=0$ this final operation is not necessary, and the particles' updating in the first step reduces to basic \textsc{sisr}. Values of $k$ in $\{1;  \ldots;n-1 \}$ induce, instead, a lookahead structure in which at the current time $t$ the optimal proposal for the delayed particles leverages information of response data $ \by_{t-k:t}$ that are not only  contemporaneous to $\bz_{t-k}$, i.e., $\by_{t-k}$, but also {\em future}, namely $\by_{t-k+1}, \ldots, \by_{t}$. In this way, the samples from the sub-trajectory $\bz_{1:t-k \mid t}$ of $\bz_{1:t \mid t}$ at time $t$ are more compatible  with the sampling density $p(\bz_{1:t}\mid \by_{1:t})$ of interest and hence, when completed with the last $k$ terms drawn from $p(\bz_{t-k+1:t} \mid \bz_{1:t-k}=\bz_{1:t-k \mid t}, \by_{t-k+1:t})$,  produce a sequential sampling scheme  from $p(\bz_{1:t}\mid \by_{1:t})$ with improved mixing and reduced degeneracy issues relative to basic \textsc{sisr}. Although the magnitude of such gains clearly grows with $k$, as illustrated in  Sect.~\ref{sec.5}, setting $k=1$ already provides empirical evidence of improved performance relative to basic \textsc{sisr}, without major computational costs.

To implement the aforementioned strategy it is first necessary to ensure that the lookahead proposal belongs to a class of random variables which allow efficient sampling, while having a tractable closed-form expression for the associated importance weights.  Proposition~\ref{prop2} shows that this is the case under model \eqref{eq3}--\eqref{eq5}.

\begin{Proposition} 
	\label{prop2}
	Under the augmented model in \eqref{eq3}--\eqref{eq5}, the  lookahead proposal mentioned above has the form
	\begin{equation}
	\begin{split}
	\label{eq17}
	&p(\bz_{t-k}\mid \bz_{1:t-k-1},\by_{t-k:t})\\
	&\quad\quad = \int p(\bz_{t-k:t}\mid \bz_{1:t-k-1},\by_{t-k:t}) d\bz_{t-k+1:t},
	\end{split}
	\end{equation}
	where $p(\bz_{t-k:t}\mid \bz_{1:t-k-1}{,}\by_{t-k:t})$ is the density of a  truncated normal  ${\textsc{tn}_{m(k{+}1)}}{(}\br_{{t-k:t\mid t{-}k{-}1}}{,}\bS_{t-k:t\mid t-k-1}{;}\mathbb{A}_{\by_{t-k:t}}{)}$ with parameters $\br_{t-k:t\mid t-k-1}=  \mathbb{E}(\bz_{t-k:t}\mid \bz_{1:t-k-1})$ and $\bS_{t-k:t\mid t-k-1}=  \mbox{\normalfont var}(\bz_{t-k:t}\mid \bz_{1:t-k-1})$. The importance weights $w_t=w(\bz_{1:t-k})$ are, instead, proportional to
	\begin{equation}
		\label{eq18}
			\dfrac{p(\by_{t-k:t}\mid \bz_{1:t-k-1})}
			{p(\by_{t-k:t-1}\mid \bz_{1:t-k-1})}= \dfrac{\Phi_{m (k+1)}(\bmu_t;\bSigma_t) }
			{\Phi_{m k}(\bar{\bmu}_t;\bar{\bSigma}_t) },
	\end{equation}
	where the mean vectors are  $\bmu_t = \bB_{t-k:t} \br_{t-k:t\mid t-k-1}$ and $\bar{\bmu}_t = \bB_{t-k:t-1}\br_{t-k:t-1\mid t-k-1}$, whereas the covariance matrices are defined as $\bSigma_t = \bB_{t-k:t} \bS_{t-k:t\mid t-k-1} \bB_{t-k:t}$  and $\bar{\bSigma}_t = \bB_{t-k:t-1} \bS_{t-k:t-1\mid t-k-1} \bB_{t-k:t-1}$.
\end{Proposition}

\begin{algorithm*}[t]
	\caption{\normalsize \fontfamily{lmss}\selectfont  Lookahead particle filter to draw from $p(\btheta_{t} \mid \by_{1:t})$, for $t=1, \ldots, n$ [\textsc{auf} version with \textsc{kf} steps]} 
	\label{algo4}
	\fontfamily{lmss}\selectfont 
	\vspace{2pt}
	Set $k$, and initialize $\ba_{0\mid 0}^{(r)} = \ba_0$ for $r=1, \ldots, R$ and $\bP_{0\mid0}=\bP_0$.\\
	\For(){$t$ \mbox{from} $1$ to $k$}
	{
	\vspace{2pt}
	 {\bf [1]}	Sample $\btheta^{(1)}_{t\mid t}, \ldots, \btheta^{(R)}_{t\mid t}$ from Algorithm \ref{algo1} [this can be done efficiently in an exact manner since $k$ is usually small].
	}
	\For(){$t$ \mbox{from} $k+1$ to $n$}
	{	\vspace{3pt}
		{\bf [2]} Define the vectors and matrices that are required to perform steps {\bf [3]} and {\bf [4]}.
		\vspace{-1pt}
		\begin{description}
		\item  \ {\bf [2.1]}  Set $\bP_{t-k\mid t-k-1} = \bG_{t-k}\bP_{t-k-1\mid t-k-1}\bG_{t-k}^\intercal + \bW_{t-k}$ [\textsc{kf}] and compute $\bS_{t-k:t\mid t-k-1}$ as in Sect.~\ref{sec.422}.\\
		\vspace{1pt}
		\item  \ {\bf [2.2]}  Set $\bP_{t-k\mid t-k}= \bP_{t-k\mid t-k-1} - \bP_{t-k\mid t-k-1}\bF_{t-k}^\intercal \bS_{t-k\mid t-k-1}^{-1}\bF_{t-k} \bP_{t-k\mid t-k-1}$ [\textsc{kf}] .\\
		\vspace{-1pt}
		\item   \ {\bf [2.3]} For $r=1, \ldots, R$, set $\ba_{t-k\mid t-k-1}^{(r)} = \bG_{t-k}\ba_{t-k-1\mid t-k-1}^{(r)}$ [\textsc{kf}]  and compute ${\br}_{t-k:t\mid t-k-1}^{(r)}$ as in Sect.~\ref{sec.422}.
		\end{description}
				\vspace{0pt}
{\bf [3]} Implement the resampling step under the \textsc{auf} version.		
		\begin{description}
				\vspace{-1pt}
\item \  {\bf [3.1]}  For $r=1, \ldots, R$, calculate the  importance weight $w^{(r)}_{t}$ via \eqref{eq18}.\\
\item \   {\bf [3.2]} Sample $(\bar{\ba}_{t-k\mid t-k-1}^{(1)},\bar{\br}_{t-k:t\mid t-k-1}^{(1)}),\ldots,(\bar{\ba}_{t-k\mid t-k-1}^{(R)},\bar{\br}_{t-k:t\mid t-k-1}^{(R)})$  from $\sum_{r=1}^{R} w_t^{(r)}\delta_{({\ba}_{t-k \mid t-k-1}^{(r)},{\br}_{t-k:t\mid t-k-1}^{(r)})}$.
			\end{description}
							\vspace{-3pt}
		\For(){$r$ \mbox{from} $1$ to $R$}{
				\vspace{3pt}
						{\bf [4]} Update the delayed particle $\bz^{(r)}_{t-k\mid t}$ and sample $\btheta^{(r)}_{t \mid t}$.
						\vspace{-3pt}
			\begin{description}
				\item {\bf [4.1]} Sample $(\bz^{(r)\intercal}_{t-k\mid t},\bar{\bz}_{t-k+1:t \mid t}^{(r)\intercal})^\intercal$ from a $\textsc{tn}_{m(k+1)}(\bar{\br}_{t-k:t\mid t-k-1},\bS_{t-k:t\mid t-k-1};\mathbb{A}_{\by_{t-k:t}})$.\\
				\vspace{0pt}
				\item {\bf [4.2]} Set $\ba_{t-k\mid t-k}^{(r)} = \bar{\ba}_{t-k\mid t-k-1}^{(r)} + \bP_{t-k\mid t-k-1}\bF_{t-k}^\intercal \bS_{t-k\mid t-k-1}^{-1}(\bz^{(r)}_{t-k\mid t} - \bar{\br}_{t-k\mid t-k-1}^{(r)})$  [\textsc{kf}].\\
				\vspace{0pt}
				\item {\bf [4.3]} Compute $\ba_{t\mid t}^{*(r)}$ and $\bP_{t\mid t}^{*(r)}$ by performing $k$ recursions of the \textsc{kf} updates applied to \eqref{eq4}--\eqref{eq5} from \\ 
\vspace{-3pt}
\quad  \ $t-k+1$ to $t$ with observations $\bz_{t-k+1:t}=\bar{\bz}_{t-k+1:t \mid t}^{(r)}$ and starting moments $\ba_{t-k\mid t-k}^{(r)} $ and $\bP_{t-k\mid t-k}$.\\
				\item {\bf [4.4]}  Sample $\btheta^{(r)}_{t\mid t}$ from the $ \mbox{N}_p(\ba_{t\mid t}^{*(r)}, \bP_{t\mid t}^{*(r)})$.
			\end{description}
		\vspace*{-10pt}
		}
	}
\end{algorithm*}

To complete the procedure for sampling from $p(\bz_{1:t}\mid \by_{1:t})$ we further require $p(\bz_{t-k+1:t}\mid \bz_{1:t-k}, \by_{t-k+1:t})$. As clarified in Proposition \ref{prop3}, also such a quantity is the density of a multivariate truncated normal.
\begin{Proposition}
\label{prop3}
Under model \eqref{eq3}--\eqref{eq5}, it holds
\begin{align}
&(\bz_{t-k+1:t}\mid \bz_{1:t-k}, \by_{t-k+1:t}) \label{eq19}\\
&\sim \textsc{tn}_{m k}(\br_{t-k+1:t\mid t-k},
\bS_{t-k+1:t\mid t-k};\mathbb{A}_{\by_{t-k+1:t}}), \nonumber
\end{align}
with parameters $\br_{t-k+1:t\mid t-k}= \mathbb{E}(\bz_{t-k+1:t}\mid \bz_{1:t-k})$ and $\bS_{t-k+1:t\mid t-k}=  \mbox{\normalfont var}(\bz_{t-k+1:t}\mid \bz_{1:t-k})$.
\end{Proposition}

Note that the expression of the importance weights in equation \eqref{eq18} does not depend on $\bz_{t-k}$, and, hence, also in this case the resampling step can be performed before sampling from \eqref{eq17}, thus leading to an \textsc{auf} routine. Besides improving efficiency, such a strategy allows to combine the particle generation in \eqref{eq17} and  the completion of the last $k$ terms of $\bz_{1:t \mid t}$ in \eqref{eq19} within a single sampling step from the joint $[m(k+1)]$-variate truncated normal distribution for $(\bz_{t-k:t}\mid \bz_{1:t-k-1},\by_{t-k:t})$ reported in Proposition \ref{prop2}. The first $m$-dimensional component of this vector yields the new delayed particle for $\bz_{t-k \mid t}$ from  \eqref{eq17}, whereas the whole sub-trajectory provides the desired sample from $p(\bz_{t-k:t}\mid \bz_{1:t-k-1},\by_{t-k:t})$ which is joined to the previously resampled particles for $\bz_{1:t-k-1 \mid t}$ to form a realization of $\bz_{1:t\mid t}$ from $p(\bz_{1:t} \mid \by_{1:t})$. Once this sample  is available, one can obtain a draw of $\btheta_{t \mid t}$ from the filtering density $p(\btheta_t\mid \by_{1:t})$ of interest by exploiting the additive representation of the \textsc{sun} and the analogy between $\bU_{1 \ 1:t\mid t}$ and $\bz_{1:t|t}$ in \eqref{eq16}. In practice, as clarified in Algorithm \ref{algo4}, the updating of $\bU_{1 \ 1:t\mid t}$ via lookahead recursion on $\bz_{1:t|t}$ and the exact sampling from the  Gaussian component of the \textsc{sun} filtering distribution for $\btheta_t$ can be effectively combined in a single online routine based on Kalman filter steps.

\vspace{-1pt}

To clarify Algorithm \ref{algo4}, note that $p(\btheta_t\mid \bz_{1:t})$ is the filtering density of  the Gaussian dynamic linear model defined in  \eqref{eq4}--\eqref{eq5}, for which the Kalman filter can be directly implemented,  once the trajectory $\bz_{1:t \mid t}$ has been generated from $p(\bz_{1:t}\mid \by_{1:t})$ via the lookahead filter. Let $\ba_{t-k-1\mid t-k-1} = \mathbb{E}(\btheta_{t-k-1}\mid \bz_{1:t-k-1})$, $\bP_{t-k-1\mid t-k-1} = \textnormal{var}(\btheta_{t-k-1}{\mid} \bz_{1:t-k-1})$ and  $\ba_{t-k{\mid} t-k-1} {=} \ \mathbb{E}(\btheta_{t-k}{\mid} \bz_{1:t-k-1})$, $\bP_{t-k\mid t-k-1} = \textnormal{var}(\btheta_{t-k}\mid \bz_{1:t-k-1})$ be the mean vector and covariance matrices for the Gaussian filtering and predictive distributions produced by the standard Kalman filter recursions at time $t-k-1$ under model  \eqref{eq4}--\eqref{eq5}. Besides being necessary to draw values from the states' filtering and predictive distributions, conditioned on the trajectories of $\bz_{1:t \mid t}$ sampled from $p(\bz_{1:t}\mid \by_{1:t})$, such quantities are also sufficient to update online the lookahead parameters $\br_{t-k:t\mid t-k-1}$ and $\bS_{t-k:t\mid t-k-1}$ that are required to compute the importance weights in Proposition \ref{prop2}, and to sample from the $[m(k+1)]$-variate truncated normal density $p(\bz_{t-k:t}\mid \bz_{1:t-k-1},\by_{t-k:t})$ under the auxiliary filter. In particular, the formulation of the dynamic model in \eqref{eq4}--\eqref{eq5} implies that $\br_{t-k:t\mid t-k-1}= \mathbb{E}(\bz_{t-k:t}\mid \bz_{1:t-k-1})=\mathbb{E}(\bF_{t-k:t}\btheta_{t-k:t}\mid \bz_{1:t-k-1})$, and, therefore, $\br_{t-k:t\mid t-k-1}$ can be expressed as a function of $\ba_{t-k\mid t-k-1}$ via the direct application of the law of the iterated expectations by stacking the $m$-dimensional vectors $\bF_{t-k}\ba_{t-k\mid t-k-1}$, \ $\bF_{t-k+1}\bG_{t-k+1}\ba_{t-k\mid t-k-1}, \ \ldots,$ \ $\bF_{t}\bG_{t-k+1}^{t} \ba_{t-k\mid t-k-1}$, where $\bG_{l}^s$ is defined as in Sect.~\ref{subsec:Smoothing}.

A similar reasoning can be applied to write the  covariance matrix $\bS_{t-k:t\mid t-k-1}=\mbox{\normalfont var}(\bz_{t-k:t}\mid \bz_{1:t-k-1})$ as a function of $\bP_{t-k\mid t-k-1}$. In particular letting $l_{-}=l-1$, the  $m \times m$  diagonal blocks of  $\bS_{t-k:t\mid t-k-1}$ can obtained sequentially after noticing that 
\begin{align*}
&\bS_{t-k:t\mid t-k-1[ll]}=\mbox{var}(\bz_{t-k+l_{-}}\mid \bz_{1:t-k-1})\\
&=\bF_{t-k+l_{-}}\bP_{t-k+l_{-}\mid t-k-1}\bF^{\intercal}_{t-k+l_{-}}+\bV_{t-k+l_{-}},
\end{align*}
for every $l=1, \ldots, k+1$, where the states' covariance matrix $\bP_{t-k+l_{-}\mid t-k-1}$ at time $t-k+l_{-}$ can be expressed as a function of $\bP_{t-k\mid t-k-1}$ via the recursive equations $\bP_{t-k+l_{-}\mid t-k-1}=\bG_{t-k+l_{-}}\bP_{t-k+l_{-}-1\mid t-k-1}\bG^{\intercal}_{t-k+l_{-}}+\bW_{t-k+l_{-}}$, for every $l=2, \ldots, k+1$. Moreover, letting $l_{-}=l-1$ and $s_{-}=s-1$, also the  off-diagonal blocks can be obtained in a related manner, after noticing that the generic block of $\bS_{t-k:t\mid t-k-1}$ is defined as
\vspace{-5pt}
\begin{align*}
&\bS_{t-k:t\mid t-k-1[sl]}=\bS^{\intercal}_{t-k:t\mid t-k-1[ls]}\\
&=\mbox{cov}(\bF_{t-k+s_{-}}\btheta_{t-k+s_{-}},\bF_{t-k+l_{-}} \btheta_{t-k+l_{-}}\mid \bz_{1:t-k-1})\\
&=\bF_{t-k+s_{-}}\bG_{t-k+l}^{t-k+s_{-}}\bP_{t-k+l_{-}\mid t-k-1}\bF_{t-k+l_{-}}^\intercal,
\end{align*}
for every $s=2, \ldots, k+1$ and $l=1, \ldots, s-1$,  where the matrix $\bP_{t-k+l_{-}\mid t-k-1}$ can be expressed as a function of $\bP_{t-k\mid t-k-1}$ via the recursive equations reported above.

According to these results, the  partially collapsed lookahead particle filter for sampling recursively from $p(\btheta_t \mid \by_{1:t})$ simply requires to store and update, for each particle trajectory, the sufficient statistics $\ba_{t-k\mid t-k-1}$ and $\bP_{t-k\mid t-k-1}$ via Kalman filter recursions applied to model \eqref{eq4}--\eqref{eq5}, with every $\bz_t$ replaced by the particles generated under the lookahead routine. As previously discussed, also this updating requires only the moments $\ba_{t-k\mid t-k-1}$ and $\bP_{t-k\mid t-k-1}$ computed recursively as a function of the delayed particles' trajectories. This yields to a computational complexity per iteration that is constant with time, as it does not require to compute quantities whose dimension grows with $t$. In addition, as discussed in Remark \ref{rem1}, such a dual interpretation combined with our \textsc{sun} closed-form results, provides novel theoretical support to the Rao--Blackwellized particle filter introduced by \citet{andrieu2002}.

\begin{Remark}
	\label{rem1}
	The Rao--Blackwellized particle filter  by \citet{andrieu2002} for $p(\btheta_t\mid \by_{1:t})$ can be directly obtained as a special case of Algorithm \ref{algo4}, setting $k=0$.
\end{Remark}

Consistent with Remark \ref{rem1}, the Rao--Blackwellized idea \citep{andrieu2002} actually coincides with a partially collapsed filter which only updates, without lookahead strategies, the truncated normal component in the \textsc{sun} additive representation of the states' filtering distribution, while maintaining the Gaussian term exact. Hence, although this method was originally motivated, in the context of dynamic probit models, also by the apparent lack of an ``optimal" closed-form \textsc{sisr} for the states' filtering distribution, our results  actually show that such a strategy is expected to yield improved performance relative to the ``optimal" particle filter for sampling directly from $p(\btheta_t \mid \by_{1:t})$. In fact, unlike this filter, which is actually available according to  Sect.~\ref{sec.421}, the Rao--Blackwellized idea avoids the unnecessary autocorrelation in the Gaussian component of the \textsc{sun} representation, and relies on an optimal particle filter for the multivariate truncated normal part. In addition, Remark \ref{rem1} and the derivation of the whole  class of partially collapsed lookahead filters suggest that setting $k>0$ is expected to yield further gains relative to the  Rao--Blackwellized particle filter; see Sect.~\ref{sec.5} for quantitative evidence supporting these results.

\vspace{-6pt}
\section{Illustration on financial time series}
\label{sec.5}
\vspace{-5pt}
Recalling Sects.~\ref{sec_1}--\ref{sec.4}, our core contribution in this article  is not on developing  innovative dynamic models for binary data with  improved  ability in recovering some ground-truth generative process, but on providing novel closed-form expressions for the filtering, predictive and smoothing distributions under a broad class of routine-use dynamic probit models, along with new Monte Carlo and sequential Monte Carlo strategies for accurate learning of such distributions and the associated functionals in practical applications.
\begin{figure*}[t]
	\centering
	\includegraphics[width=1\linewidth,height=6.7cm]{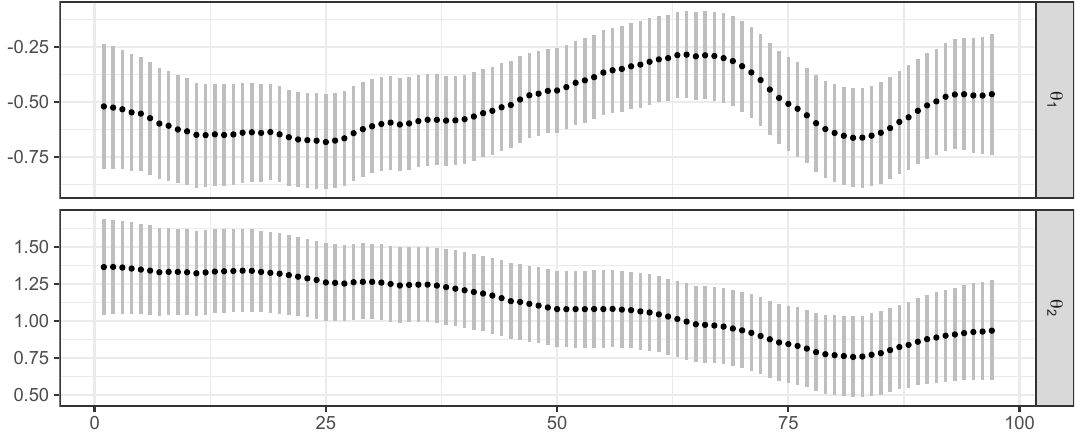}
	\caption{\footnotesize{\fontfamily{lmss}\selectfont  Pointwise median and interquartile range for the smoothing distributions of $\theta_{1t}$ and $\theta_{2t}$ in model \eqref{eq12},  for the time window from  January 4th, 2018 to May 31st, 2018. The quartiles are computed from $10^5$ samples produced by Algorithm \ref{algo1}.}}
	\label{fig:smoothing}
\end{figure*}

Consistent with the above discussion, we illustrate the practical utility of the closed-form results for the filtering, predictive and smoothing distributions derived in Sect.~\ref{sec.3} directly on a realistic real-world dataset, and assess the performance gains of the Monte Carlo strategies developed in Sect.~\ref{sec.4}. The  focus will be on the accuracy in recovering the whole exact \textsc{sun} distributions of interest, and not just pre-selected functionals. In fact, accurate learning of the entire exact distribution is more challenging and implies, as a direct consequence, accuracy in approximating the associated exact functionals. These assessments are illustrated with a focus on a realistic financial application considering a dynamic probit regression for the daily opening directions of the  French \textsc{cac}40 stock market index from January 4th, 2018 to March 29th, 2019. In this study, the variable $y_t$ is defined on a binary scale, with $y_t=1$ if the opening value of the \textsc{cac}40 on day $t$ is greater than the corresponding closing value in the previous day, and $y_t=0$ otherwise. Financial applications of this type have been a source of particular interest in past and recent years \citep[e.g.,][]{kim_2000,kara_2011,atkins_2018}, with common approaches combining a wide variety of technical indicators and news information to forecast stock markets directions via complex machine  learning methods. Here, we show how a similar predictive performance can be obtained via a simple and interpretable dynamic probit regression for $y_t$, which combines past information on the opening directions of  \textsc{cac}40 with those of the \textsc{nikkei}225, regarded as binary covariates $x_t$ with dynamic coefficients. Since the Japanese market opens before the French one,  $x_t$ is available prior to $y_t$ and, hence, provides a valid predictor for each day $t$.

\vspace{-2pt}

Recalling the above discussion and leveraging the default model specifications in these settings \citep[e.g.,][]{soyer2013bayesian}, we rely on a dynamic probit regression for $y_t$ with two independent random walk processes for the  coefficients $\btheta_t=(\theta_{1t}, \theta_{2t})^{\intercal}$.  Letting $\bF_t=(1,x_t)$ and $\mbox{pr}(y_{t}=1 \mid \btheta_{t})=\Phi(\theta_{1t}+\theta_{2t}x_t;1)$, such a model can be expressed as in equations \eqref{eq1}--\eqref{eq2} via
\vspace{-2pt}
\begin{eqnarray}
\begin{split}
&p(y_{t} \mid \btheta_{t})= \Phi[(2y_t-1)\bF_t\btheta_t;1],\\ 
&\btheta_t=\btheta_{t-1}+\bvarepsilon_t, \quad \bvarepsilon_t\stackrel{\mbox{\scriptsize i.i.d.}}{\sim}  \mbox{N}_2({\bf 0}, \bW), \quad t= 1, \ldots n, 
\label{eq12}
\end{split}
\end{eqnarray}
\vspace{-10pt}

\noindent where $\btheta_0 \sim \mbox{N}_2(\ba_0, \bP_0)$, whereas $\bW$ is a time-invariant diagonal matrix. In \eqref{eq12}, the element $\theta_{1t}$ of $\btheta_t$ measures the trend in the directions of  the \textsc{cac}40 when the  \textsc{nikkei}225 has a negative opening on day $t$, whereas $\theta_{2t}$ characterizes the shift in such a trend if the opening of the   \textsc{nikkei}225 index is positive, thereby providing an interpretable probit model  with dynamic coefficients.

To evaluate performance in smoothing, filtering and prediction, we split the time window in two parts. Observations from January 4th, 2018 to May 31st, 2018 are used as batch data to study the smoothing distribution and to compare the particle filters developed in Sect.~\ref{sec.42} with other relevant competitors. In the subsequent time window, spanning from June 1st, 2018 to March 29th, 2019, the focus is instead on illustrating performance in online filtering and prediction for streaming data via the lookahead routine derived in Sect.~\ref{sec.422} --- which yields the highest approximation accuracy among the online filters evaluated in the first time window.

Figure \ref{fig:smoothing} shows the pointwise median and interquartile range of the smoothing distribution for $\theta_{1t}$ and $\theta_{2t}$, $t=1, \ldots, 97$, based on $R=10^5$ samples from Algorithm \ref{algo1}. To implement this routine, we set $\ba_0=(0,0)^{\intercal}$ and $\bP_0=\mbox{diag}(3,3)$ following the guidelines in \citet{gelman_2008} and \citet{chopin_2017} for probit regression. The errors' variances in the diagonal matrix $\bW$ are instead set equal to $0.01$  as suggested by a graphical search of the maximum for the marginal likelihood computed under different combinations of $(\mbox{W}_{11},\mbox{W}_{22})$ via the analytical formula in Corollary \ref{cor:MarginalLikelihood}.

\begin{figure*}[t]
	\centering
	\includegraphics[width=1.05\linewidth,height=7.2cm]{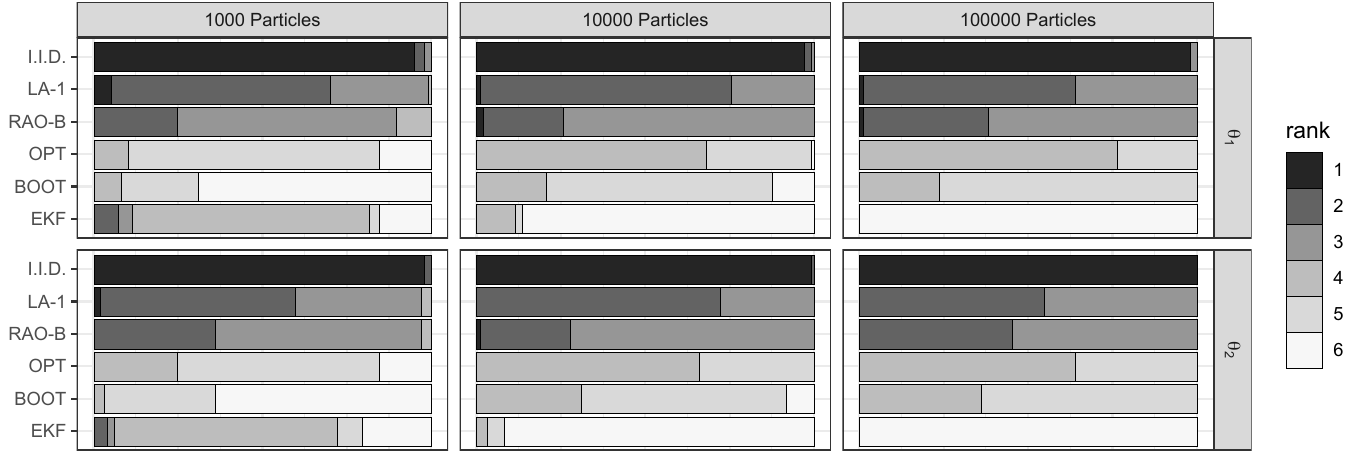}
	\caption{\footnotesize{\fontfamily{lmss}\selectfont  For the states $\theta_{1t}$ and $\theta_{2t}$, barplots representing the relative frequencies of  global rankings for the six sampling schemes, in terms of accuracy in approximating the exact \textsc{sun} filtering distributions over the time window analyzed. For each scheme and time $t=1, \ldots, 97$, the  accuracy is measured via the median Wasserstein distance  (over $100$ replicated experiments) between the empirical filtering distribution computed from $10^3,10^4$ and $10^5$ particles, respectively, and the one  obtained by direct evaluation of the associated exact density from \eqref{eq10} on two grids of $2000$ equally spaced values for $\theta_{1t}$ and $\theta_{2t}$. This allows to compute, for every $t=1, \ldots, 97$, the ranking of each sampling scheme in terms of accuracy in approximating the exact filtering density at time $t$, and to derive the associated barplot summarizing the distribution of the rankings over the whole window.  }}
	\label{f3}
\end{figure*}

As shown in Fig.~\ref{fig:smoothing}, the dynamic states  $\theta_{1t}$ and $\theta_{2t}$ tend to concentrate around negative and positive values, respectively, for the entire smoothing window, thus highlighting a general concordance between the opening patterns of  \textsc{cac}40 and \textsc{nikkei}225. However, the strength of this association varies in time, supporting our proposed dynamic probit over static specifications. For example, it is possible to observe a decay in  $\theta_{1t}$ and $\theta_{2t}$ on April--May, 2018 which reduces the association among \textsc{cac}40 and \textsc{nikkei}225, while inducing a general negative trend for the opening directions of the French market. This could be due to the overall instability in the Eurozone on  April--May, 2018 caused by the uncertainty after the Italian and British elections during those months.

To clarify the computational improvements of the methods developed in Sects.~\ref{sec.41} and \ref{sec.42}, we also compare, in Fig.~\ref{f3} and in Table \ref{table1}, their  performance against the competing strategies mentioned in Sect.~\ref{sec_1}. Here, the focus is on the accuracy and computational cost in approximating the exact filtering distribution at time $t=1,\ldots,97$, thereby allowing the implementation of the filters discussed in Sect.\ \ref{sec_1}. The competing methods include the extended Kalman filter \citep{uhlmann1992} (\textsc{ekf}), the bootstrap particle filter \citep{gordon1993} (\textsc{boot}), and the Rao--Blackwellized  (\textsc{rao-b}) sequential Monte Carlo  strategy by \citet{andrieu2002}, which has been discussed in Sect.~\ref{sec.422} and exploits the hierarchical representation \eqref{eq3}--\eqref{eq5} of model \eqref{eq1}--\eqref{eq2}. Although being a popular solution in routine implementations, the extended Kalman filter relies on a quadratic approximation of the probit log-likelihood which leads to Gaussian filtering distributions, thereby affecting the quality of online learning when imbalances in the data induce skewness. The bootstrap particle filter \citep{gordon1993} provides, instead, a general \textsc{sisr} that relies on the importance density $p(\btheta_t \mid \btheta_{t-1})$, thus failing to account effectively for  information in $\by_t$, when proposing particles.  Rao--Blackwellized sequential Monte Carlo \citep{andrieu2002} aims at providing an alternative particle filter, which also addresses the apparent unavailability of an analytic form for  the ``optimal"  particle filter \citep{doucet2000}. The authors overcome this issue by proposing a sequential Monte Carlo strategy for the Rao--Blackwellized  density $p(\bz_{1:t} \mid \by_{1:t})$ of the partially observed Gaussian responses $\bz_{1:t}$ in model \eqref{eq3}--\eqref{eq5} and compute, for each  trajectory $\bz_{1:t \mid t}$, relevant moments of $(\btheta_t \mid \bz_{1:t\mid t})$ via classical Kalman filter updates --- applied to model \eqref{eq4}--\eqref{eq5} --- which are then averaged across the particles to obtain Monte Carlo estimates for the moments of $(\btheta_t \mid \by_{1:t})$. As specified in Remark \ref{rem1}, this solution, when adapted to draw samples from $p(\btheta_t\mid \by_{1:t})$, is a special case of the sequential strategy  in Sect.~\ref{sec.422}, with no lookahead, i.e., $k=0$.

Although the above methods yield state-of-the-art solutions, the proposed strategies are motivated by the apparent absence of a closed-form filter for \eqref{eq1}--\eqref{eq2}, that is, in fact, available according to our findings in Sect.~\ref{sec.3}. Consistent with this argument, we evaluate the accuracy of \textsc{efk}, \textsc{boot} and \textsc{rao-b} in approximating the exact filtering distribution obtained, for each $t=1, \ldots, 97$, via direct evaluation of the density from \eqref{eq10}. These performances are also compared with those of the new methods  proposed in Sect.~\ref{sec.4}. These include the filtering version of the  i.i.d.\ sampler (\textsc{i.i.d.}) in Sect.~\ref{sec.41}, along with the  ``optimal'' particle filter  (\textsc{opt}) presented in Sect.~\ref{sec.421},  and the lookahead sequential Monte Carlo routine derived in Sect.~\ref{sec.422}, setting $k=1$ (\textsc{la-1}).

\renewcommand{\arraystretch}{1.12}
\begin{table*}[t]
	\centering
	\begin{tabular}{lcccccc}
			\hline
	 \qquad  \qquad 	& \multicolumn{6}{c}{\textsc{accuracy}}\\
		\hline 
		 &$\theta_{1t}$ $[R=10^3]$&$\theta_{2t}$ $[R=10^3]$\quad &\quad  $\theta_{1t}$ $[R=10^4]$&$\theta_{2t}$ $[R=10^4]$\quad &\quad $\theta_{1t}$ $[R=10^5]$&$\theta_{2t}$ $[R=10^5]$\\ 
\hline
		\textsc{i.i.d.} &0.01917 [{\bf 1}]&0.02362  [{\bf 1}]&0.00606  [{\bf 1}]&0.00748  [{\bf 1}]& 0.00199 [{\bf 1}]&0.00245  [{\bf 1}]\\ 
		\textsc{la--1} &0.02558  [{\bf 2}] &0.03588 [{\bf 2}] &0.00838  [{\bf 2}] &0.01133  [{\bf 2}] &0.00273  [{\bf 2}] &0.00379  [{\bf 2}] \\
		\textsc{rao--b} &0.02700 [{\bf 3}] &0.03700  [{\bf 3}] &0.00885 [{\bf 3}] &0.01201 [{\bf 3}] &0.00278 [{\bf 3}] &0.00383 [{\bf 3}] \\
				\textsc{opt} &0.06642 [{\bf 5}]&0.09063 [{\bf 4}]&0.02196   [{\bf 4}]&0.03077   [{\bf 4}]&0.00687   [{\bf 4}]&0.00958   [{\bf 4}]\\
		\textsc{boot} &0.07237 [{\bf 6}]&0.10021 [{\bf 5}]&0.02325  [{\bf 5}]&0.03225  [{\bf 5}]&0.00728  [{\bf 5}]&0.00992  [{\bf 5}]\\
		\textsc{ekf} &0.06108 [{\bf 4}] &0.10036  [{\bf 6}] &0.05853  [{\bf 6}]&0.09824  [{\bf 6}]&0.05829  [{\bf 6}]&0.09802  [{\bf 6}]\\
		\hline
		 \qquad \qquad	& \multicolumn{6}{c}{\textsc{computational cost}}\\
		 		\hline
		 \textsc{i.i.d.} &\multicolumn{6}{l}{$\mathcal{O}(tp^3 + t^3 m^3 + R[p^2 + t^2m^2C(mt)])$}\\ 
		\textsc{la--1} &\multicolumn{6}{l}{$\mathcal{O}(t(p^3 + m^3) + tR[p^2 + pm + m^2C(2m)]+ tM[m^2 + Rm])$}\\
		\textsc{rao--b} &\multicolumn{6}{l}{$\mathcal{O}(t(p^3+m^3) +tR[p^2 + pm + m^2C(m)] + tM[m^2 + Rm])$}\\
				\textsc{opt} &\multicolumn{6}{l}{$\mathcal{O}(t(p^3+m^3) +tR[p^2 + pm + m^2C(m)] + tM[m^2 + Rm])$}\\
		\textsc{boot} &\multicolumn{6}{l}{$\mathcal{O}(t(p^3+m^3) + tR(p^2+pm) +tM[m^2 + Rm])$}\\
		\textsc{ekf} &\multicolumn{6}{l}{$\mathcal{O}(t[p^3+m^3 + Mm^2])$}\\
		\hline
	\end{tabular}
	\caption{\footnotesize{\fontfamily{lmss}\selectfont  For the  states $\theta_{1t}$ and $\theta_{2t}$, averaged accuracy in approximating the exact \textsc{sun} filtering distribution at $t=1, \ldots, 97$, and computational cost for obtaining a sample of dimension $R$ from such a filtering distribution at time $t$. For each scheme, the accuracy is measured via the Wasserstein distance between the empirical filtering distribution computed from $10^3,10^4$ and $10^5$ particles, respectively, and the one  obtained via direct evaluation of the associated exact \textsc{sun} density from \eqref{eq10} on two grids of $2000$ equally spaced values for $\theta_{1t}$ and $\theta_{2t}$. For each $t$, we first compute the median Wasserstein distance from $100$ replicated experiments, and then average such quantities across time. Numbers in square brackets denote the ranking in each column. The costs are derived for the case in which the importance weights are evaluated via Monte Carlo based on $M$ samples. For the  \textsc{ekf}, we provide the cost of  the \textsc{kf} recursions, when the probit likelihood is evaluated via $M$ Monte Carlo samples.
	} 
	\label{table1}}
\end{table*}

For the two dynamic state variables $\theta_{1t}$ and $\theta_{2t}$, the accuracy of each sampling scheme is measured via the Wasserstein distance \citep[e.g.,][]{villani_2008} between the empirical filtering distribution computed, for every time $t=1, \ldots, 97$, from $R=10^3$, $R=10^4$ and $R=10^5$ particles produced by that specific scheme and the one  obtained via the direct evaluation of the associated exact density from \eqref{eq10} on two grids of $2000$ equally spaced values for $\theta_{1t}$ and $\theta_{2t}$.  For the sake of clarity, with a little abuse of terminology, the term {\em particle} refers both to the samples of the sequential Monte Carlo methods and to those obtained under i.i.d.\ sampling from the \textsc{sun}. The Wasserstein distance is computed via the \texttt{R} function \texttt{wasserstein1d}. Note also that, although \textsc{ekf} and \textsc{rao-b} focus, mostly, on moments of $(\btheta_t \mid \by_{1:t})$, such strategies can be adapted to sample from an approximation of the filtering distribution. Figure \ref{f3} displays, for the two states and for varying number of particles, the frequencies of the global rankings of the different schemes, out of the $97$ time instants. Such rankings are computed according to the median Wasserstein distance obtained, for each $t=1,\ldots,97$, from $100$ replicated experiments. The overall averages across time of these median Wasserstein distances are reported in Table \ref{table1}, along with computational costs for obtaining $R$ samples  from the filtering at time $t$ under each scheme; see Appendix B for detailed derivations of such costs.

\begin{figure*}[t]
	\centering
	\includegraphics[width=1\linewidth,height=7.5cm]{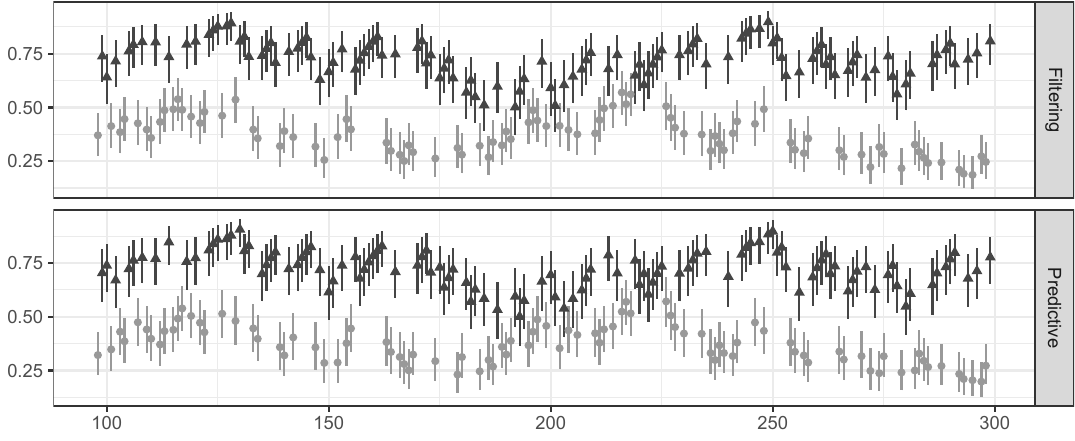}
	\caption{\footnotesize{\fontfamily{lmss}\selectfont  Median and interquartile range of the filtering and predictive distributions for $\Phi(\theta_{1t}+x_t\theta_{2t};1)$ computed from $10^5$ particles produced by the lookahead particle filter in Algorithm \ref{algo4} for the second time window. Black and grey segments denote days in which  $x_t=1$ and $x_t=0$, respectively. }
	}
	\label{f2}
\end{figure*}

Figure \ref{f3} and Table \ref{table1} confirm that the \textsc{i.i.d.} sampler  in Sect.~\ref{sec.41} over-performs the competitors in accuracy, since the averaged median Wasserstein distances from the exact filtering distribution are lower than those of the other schemes under all settings, and the ranking of the \textsc{i.i.d.}  is $1$  in almost all the $97$ times. This improved performance comes, however, with a higher computational complexity, especially in the sampling from $(m t)$-variate truncated normals in the \textsc{sun} additive representation, which yields a cost depending on $C(mt)$, i.e., the average number of proposed draws required to accept one sample. While the improved accuracy of  \textsc{i.i.d.} justifies such a cost  in small-to-moderate dimensions, as $t$ increases the \textsc{i.i.d.} becomes progressively impractical, thus motivating scalable particle filters with linear cost in $t$, such as \textsc{boot}, \textsc{rao-b}, \textsc{opt} and \textsc{la-1}. In our basic \texttt{R} implementation, we found that the proposed \textsc{i.i.d.} sampler has reasonable runtimes (of a couple of minutes) also for  larger series with $mt \approx 300$. However, in much higher dimensions the  particle filters become orders of magnitude faster and still practically effective.

As expected, the \textsc{opt}  filter  in Sect.~\ref{sec.421} tends to improve the performance of \textsc{boot}, since this strategy is optimal within the class where  \textsc{boot} is defined. However, as discussed in Sects.~\ref{sec.421} and \ref{sec.422}, both methods induce unnecessary autocorrelation in the Gaussian part of the \textsc{sun} filtering distribution, thus yielding suboptimal solutions relative to particle filters that  perform sequential Monte Carlo only on the multivariate truncated normal component. The accuracy gains of \textsc{rao-b} and \textsc{la-1} relative to \textsc{boot} and \textsc{opt} in Fig.~\ref{f3} and Table~\ref{table1} provide empirical evidence in support of this argument, while displaying additional improvements of the lookahead strategy derived in Sect.~\ref{sec.422} over \textsc{rao-b}, even when $k$ is set just  to $1$, i.e., \textsc{la-1}.  As shown in Table~\ref{table1}, the complexities of  \textsc{la-1} and \textsc{rao-b} are of the same order, except for sampling from bivariate truncated normals under \textsc{la-1} instead of univariate ones as in \textsc{rao-b}. This holds for any fixed $k$, with the additional sampling cost being $C(m[k+1])$. However, consistent with the results in Fig.~\ref{f3} and Table \ref{table1} it suffices to set $k$ quite small to already obtain some accuracy gains, thus making such increments in computational cost affordable in practice. The \textsc{ekf} is, overall, the less accurate solution since, unlike the other methods, it relies on a Gaussian approximation of the \textsc{sun} filtering distribution. This is only beneficial relative to \textsc{boot} and \textsc{opt} when the number of particles is small, due to the reduced mixing of such strategies induced by the autocorrelation in the Gaussian component of the \textsc{sun} additive representation. All these results remained consistent also when comparing other quantiles of the Wasserstein distance across experiments and when studying the accuracy in approximating pre-selected  functionals of interest.

Motivated by the accurate performance of the novel lookahead strategy  in Sect.~\ref{sec.422}, we apply \textsc{la-1} to provide scalable online filtering and prediction for model \eqref{eq12} from June 1st, 2018 to March 29th, 2019. Following the idea of sequential inference, the particles are initialized exploiting the marginal smoothing distribution of May 31, 2018 from the batch analysis. Figure \ref{f2} outlines median and interquartile range for the filtering and predictive distribution of the probability that  \textsc{cac40} has a positive opening in each day of the window considered for online inference. These two distributions can be easily obtained by applying the function  $\Phi(\theta_{1t}+x_t\theta_{2t};1)$ to the particles of the states filtering and predictive distribution. In line with Fig.~\ref{fig:smoothing}, a positive opening of the \textsc{nikkei}225 provides, in general, a high estimate for the probability that $y_t=1$, whereas a negative opening tends to favor the event $y_t=0$. However, the strength of this result evolves over time with some periods showing less evident shifts in the probabilities process when $x_t$ changes from $1$ to $0$. One-step-ahead prediction, leveraging the samples of the predictive distribution for the probability process, led to a correct classification rate of $66.34\%$ which is comparable to those obtained under more complex procedures combining a wide variety of inputs to predict stock markets directions via state-of-the-art machine  learning methods \citep[e.g.,][]{kim_2000,kara_2011,atkins_2018}.

\vspace{-8pt}

\section{Discussion}
\label{sec.6}
\vspace{-5pt}

This article shows that filtering, predictive and smoothing densities in multivariate dynamic probit models have a \textsc{sun} kernel and the associated parameters can be computed via tractable expressions. As discussed in Sects.~\ref{sec.3}--\ref{sec.5}, this result provides advances in online inference  and facilitates the implementation of tractable methods to draw i.i.d. samples from the exact filtering, predictive and smoothing distributions, thereby allowing improved Monte Carlo inference in  small-to-moderate settings. Filtering in higher dimensions can be, instead, implemented via scalable sequential Monte Carlo which exploits \textsc{sun} properties to provide novel particle filters.

Such advances motivate future research. For example, a relevant direction is to  extend the results in Sect.~\ref{sec.3} to dynamic tobit, binomial and multinomial probit models, for which closed-form filters are unavailable. In the multinomial setting a viable solution is to exploit the results in \citet{fasano2020} for the static case. Joint filtering and prediction of continuous and binary time series is also of interest \citep[][]{liu_2009}. A natural state-space model for these  data can be obtained by  allowing only the sub-vector of Gaussian variables associated with the binary data to be partially observed in \eqref{eq3}--\eqref{eq5}. However, also in this case, closed-form filters are unavailable. By combining our results  in Sect.~\ref{sec.3} with classical Kalman filter, this gap may now be covered.

As mentioned in Sects.~\ref{sec_1} and \ref{subsec:Smoothing}, estimation of possible unknown parameters characterizing the state-space model in  \eqref{eq1}--\eqref{eq2} is another relevant problem, that can be addressed by maximizing the marginal likelihood derived in Sect.~\ref{subsec:Smoothing}. This quantity can be explicitly evaluated as in Corollary \ref{cor:MarginalLikelihood} for any small-to-moderate $n$. A more scalable option in large $n$ settings is to rely on equations (62) and (66) in \citet{doucet2000} which allow to evaluate the marginal likelihood leveraging samples from particle filters. In this respect, the improved lookahead filter developed in Sect.~\ref{sec.422} is expected to yield accuracy gains also in parameter estimation, when used as a scalable strategy to evaluate marginal likelihoods. This  routine can be also adapted to sample from the joint smoothing distribution via a backward recursion. However, unlike the i.i.d. sampler in Algorithm \ref{algo1}, this approach yields an additional computational cost which is quadratic in the total number of particles $R$ \citep[e.g.,][]{doucet2000}. Since $R$ is much higher than $n$ in most applications, the i.i.d. sampler developed in Algorithm \ref{algo1} is preferable over particle smoothers in routine studies having small-to-moderate dimension, since it also yields improved accuracy by avoiding sequential Monte Carlo. Finally, additional quantitative studies beyond those  in Sect.~\ref{sec.5} can be useful for obtaining further insights on the performance of our proposed algorithms relative to state-of-the-art strategies, including recent ensemble sampling \citep{deligiannidis2020}.

\vspace{4pt}
{\footnotesize \noindent {\bf Data and Codes}. The dataset considered  in Sect.~\ref{sec.5} is available at \texttt{Yahoo Finance}. Pseudo-codes that can be easily implemented with any software are provided in Algorithms~\ref{algo1}--\ref{algo4}.}

\vspace{-15pt}

\section*{Appendix A: Proofs of the main results}
\label{app}
\vspace{-5pt}
{\bf \em Proof of Lemma \ref{lem1}}.  To prove Lemma \ref{lem1}, note that, by applying the Bayes' rule, we obtain 
\vspace{-5pt}
\begin{equation*}
p(\btheta_1 \mid \by_1) \propto p(\btheta_1)p(\by_{1} \mid \btheta_{1}),
\end{equation*}
where $p(\btheta_1)=\phi_p(\btheta_1-\bG_1\ba_0;\bG_1\bP_0\bG_1^{\intercal}+\bW_1)$ and  $p(\by_{1} \mid \btheta_{1})= \Phi_m(\bB_1\bF_1\btheta_{1}; \bB_1\bV_1\bB_1)$. The expression  for  $p(\btheta_1)$ can be obtained by noting that $\btheta_1=\bG_1\btheta_0+\bvarepsilon_1$ in \eqref{eq2}, with  $\btheta_0 \sim \mbox{N}_p(\ba_0, \bP_0)$ and $\bvarepsilon_1\sim \mbox{N}_p({\bf 0}, \bW_1)$. The form for the probability mass function of $(\by_{1} \mid \btheta_{1})$ is instead a direct consequence of equation \eqref{eq1}. Hence, combining these results and recalling \eqref{eq6}, it is clear that $p(\btheta_1 \mid \by_1)$ is proportional to the density of a \textsc{sun} with suitably--specified parameters, such that the kernel of \eqref{eq6} coincides with $\phi_p(\btheta_1-\bG_1\ba_0;\bG_1\bP_0\bG_1^{\intercal}+\bW_1)\Phi_m(\bB_1\bF_1\btheta_{1}; \bB_1\bV_1\bB_1)$. In particular, letting
\vspace{-5pt}
\begin{align*}
	\bxi_{1\mid1}&=\bG_1 \ba_0, \quad \bOmega_{1\mid1}=\bG_1\bP_0\bG_1^{\intercal}+\bW_1, \\
\bDelta_{1\mid1}&=\bar{\bOmega}_{1\mid1}{\bomega}_{1\mid1}\bF_1^{\intercal} \bB_1\bs_1^{-1}, \quad \bgamma_{1\mid1}=\bs_1^{-1}\bB_1\bF_1\bxi_{1\mid1}, \\ 
\bGamma_{1\mid1}&= \bs^{-1}_1\bB_1(\bF_1\bOmega_{1\mid1} \bF^{\intercal}_1+\bV_1)\bB_1\bs^{-1}_1,
	\end{align*}
we have that 	
\begin{align*}
	&\bgamma_{1|1}+\bDelta_{1|1}^\intercal \bar{\bOmega}_{1|1}^{-1} \bomega_{1|1}^{-1}(\btheta_1-\bxi_{1|1})\\
	&=\bs_1^{-1}\bB_1\bF_1\bxi_{1\mid1}+\bs_1^{-1}\bB_1\bF_1(\btheta_1-\bxi_{1|1})=\bs_1^{-1}\bB_1\bF_1\btheta_1,\\
& \bGamma_{1|1}{-}\bDelta_{1|1}^{\intercal}\bar{\bOmega}_{1|1}^{-1}\bDelta_{1|1}\\
&=	\bs^{-1}_1[\bB_1(\bF_1\bOmega_{1\mid1} \bF^{\intercal}_1{+}\bV_1)\bB_1-\bB_1(\bF_1\bOmega_{1\mid1} \bF^{\intercal}_1)\bB_1]\bs^{-1}_1\\
&=\bs^{-1}_1\bB_1\bV_1\bB_1\bs^{-1}_1.
	\end{align*}
with $\bs^{-1}_1$ as  in Lemma \ref{lem1}. Note that this term is introduced to make $\bGamma_{1|1}$ a correlation matrix, as required in the \textsc{sun} parametrization \citep{arellano_2006}.
Recalling  \citet{Durante2018}, and substituting these quantities in the kernel of the \textsc{sun} density \eqref{eq6}, we have
\begin{align*}
&\phi_p(\btheta_1-\bG_1\ba_0;\bG_1\bP_0\bG_1^{\intercal}+\bW_1)\\
& \quad \cdot \Phi_m(\bs^{-1}_1\bB_1\bF_1\btheta_{1}; \bs^{-1}_1\bB_1\bV_1\bB_1\bs^{-1}_1)\\
&=\phi_p(\btheta_1{-}\bG_1\ba_0;\bG_1\bP_0\bG_1^{\intercal}{+}\bW_1)\Phi_m(\bB_1\bF_1\btheta_{1}; \bB_1\bV_1\bB_1)\\
&=p(\btheta_1)p(\by_{1} \mid \btheta_{1}) \propto p(\btheta_1 \mid \by_1),
	\end{align*}
thus proving Lemma \ref{lem1}. To prove that $\bOmega_{1|1}^*$ is a correlation matrix, replace the indentity ${\bf I}_m$  with $ \bB_1\bV_1\bB_1$  in the proof of Theorem 1 by  \citet{Durante2018}. \qquad \qquad \qquad  $\qed$

\vspace{7pt}

\noindent {\bf \em Proof of Theorem \ref{teo1}}. Recalling equation  \eqref{eq2}, the proof for $p(\btheta_t \mid \by_{1:t-1})$ in \eqref{eq9} requires studying the variable $\bG_t\btheta_{t-1}+\bvarepsilon_t$, given $\by_{1:t-1}$, where 
\begin{align*}
(\btheta_{t-1} \mid \by_{1:t-1}) \sim \ & \textsc{sun}_{p,m(t-1)}(\bxi_{t-1\mid t-1}, \bOmega_{t-1\mid t-1},\\
& \qquad \quad \ \bDelta_{t-1\mid t-1}, \bgamma_{t-1\mid t-1}, \bGamma_{t-1\mid t-1}),
\end{align*}
and $\bvarepsilon_t \sim \mbox{N}_p({\bf 0}, \bW_t)$, with $\bvarepsilon_t  \perp  \by_{1:t-1}$. To address this goal, first note that, by the closure properties of the \textsc{sun} family under linear transformations \citep[][Sect.~7.1.2]{azzalini_2013}, we have that $(\bG_t\btheta_{t-1} \mid \by_{1:t-1})$ is still a \textsc{sun} with parameters $\bG_t\bxi_{t-1\mid t-1}$, $\bG_t\bOmega_{t-1\mid t-1}\bG^{\intercal}_t$, $[(\bG_t\bOmega_{t-1\mid t-1}\bG^{\intercal}_t) \odot \mbox{\bf I}_p]^{-\frac{1}{2}}\bG_t \bomega_{t-1\mid t-1} \bDelta_{t-1\mid t-1}$, $\bgamma_{t-1\mid t-1}$ and $\bGamma_{t-1\mid t-1}$. Hence, to conclude the proof of equation  \eqref{eq9}, we only need to obtain the distribution of the sum among this variable and the noise $\bvarepsilon_t \sim \mbox{N}_p({\bf 0}, \bW_t)$. This can be accomplished by considering the moment generating function of such a sum --- as done by \citet[][Sect.~7.1.2]{azzalini_2013} to prove closure under convolution. Indeed, it is straightforward to note that the product of the moment generating functions for  $\bvarepsilon_t$  and $(\bG_t\btheta_{t-1} \mid \by_{1:t-1})$ leads to the moment generating function of a \textsc{sun} random variable having parameters  $\bxi_{t\mid t-1}=\bG_t \bxi_{t-1\mid t-1}$, $\bOmega_{t\mid t-1}=\bG_t\bOmega_{t-1\mid t-1}\bG_t^{\intercal}+\bW_t$, $\bDelta_{t\mid t-1}={\bomega}^{-1}_{t\mid t-1}\bG_t {\bomega}_{t-1\mid t-1}\bDelta_{t-1\mid t-1}$, $\bgamma_{t\mid t-1}=\bgamma_{t-1\mid t-1}$ and $\bGamma_{t\mid t-1}$ $= \bGamma_{t-1\mid t-1}$. To prove \eqref{eq10} note that 
\begin{equation*}
p(\btheta_t \mid \by_{1:t}) \propto \Phi_m(\bB_t\bF_t\btheta_{t}; \bB_t\bV_t\bB_t)p(\btheta_t \mid \by_{1:t-1}) 
\end{equation*}
coincides with the posterior density in the probit model having likelihood $\Phi_m(\bB_t\bF_t\btheta_{t}; \bB_t\bV_t\bB_t)$, and \textsc{sun} prior $p(\btheta_t \mid \by_{1:t-1})$ from \eqref{eq9}. Hence,  \eqref{eq10} can be derived from Corollary 4 in \citet{Durante2018}, replacing matrix ${\bf I}_m$ in the classical probit likelihood with $ \bB_t\bV_t\bB_t$.
\qquad \qquad \qquad  $\qed$

\vspace{7pt}

\noindent {\bf \em Proof of Corollary  \ref{cor1}}. To prove Corollary  \ref{cor1}, re-write $\int \Phi_m(\bB_t\bF_t\btheta_{t}; \bB_t\bV_t\bB_t)p(\btheta_t \mid \by_{1:t-1}) \mbox{d} \btheta_t$  as 
\begin{equation*}
\frac{\int \Phi_m(\bB_t\bF_t\btheta_{t}; \bB_t\bV_t\bB_t)K(\btheta_t \mid \by_{1:t-1}) \mbox{d} \btheta_t}{\Phi_{m (t-1)}(\bgamma_{t\mid t-1}; \bGamma_{t\mid t-1})},
\end{equation*}
with $K(\btheta_t {\mid} \by_{1:t-1})=p(\btheta_t  {\mid} \by_{1:t-1})\Phi_{m (t-1)}(\bgamma_{t\mid t-1}{;} \bGamma_{t\mid t-1})$ denoting the kernel of the predictive density from \eqref{eq9}. Consistent with this result, Corollary  \ref{cor1} follows by noting that $ \Phi_m(\bB_t\bF_t\btheta_{t}; \bB_t\bV_t\bB_t)K(\btheta_t \mid \by_{1:t-1})$ is the kernel of the filtering density from  \eqref{eq10}, whose normalizing constant $\int \Phi_m(\bB_t\bF_t\btheta_{t}; \bB_t\bV_t\bB_t)K(\btheta_t \mid \by_{1:t-1}) \mbox{d} \btheta_t$ is equal to $\Phi_{m t}(\bgamma_{t\mid t}; \bGamma_{t\mid t})$. \qquad \qquad \qquad \qquad \qquad  \qquad \quad   \qquad  $\qed$

\vspace{7pt}

\noindent {\bf \em Proof of Theorem \ref{thm:JointSmoothing}}. First notice that $p(\btheta_{1:n} \mid \by_{1:n}) \propto p(\btheta_{1:n}) p(\by_{1:n} \mid \btheta_{1:n})$. Therefore, $p(\btheta_{1:n} \mid \by_{1:n})$ can be seen as the posterior density in the Bayesian model with likelihood $p(\by_{1:n} \mid \btheta_{1:n})$ and prior $ p(\btheta_{1:n}) $ for the vector $\btheta_{1:n}=(\btheta^{\intercal}_1, \ldots, \btheta^{\intercal}_n)^{\intercal}$. As  pointed out in Sect.~\ref{subsec:Smoothing}, it follows from \eqref{eq2} that $\btheta_{1:n}\sim \mbox{\normalfont N}_{p n}(\bxi,\bOmega)$, with $\bxi$ and $\bOmega$ defined in Sect.~\ref{subsec:Smoothing}. The form of $p(\by_{1:n} \mid \btheta_{1:n})$ can be obtained from  \eqref{eq1}, by noticing that $\by_1, \ldots, \by_n$ are conditionally independent given $ \btheta_{1:n}$, thus providing the joint likelihood $p(\by_{1:n} \mid \btheta_{1:n})=\prod_{s=1}^n\Phi_m(\bB_s\bF_s\btheta_{s}; \bB_s\bV_s\bB_s)$. This quantity can be re-written as $\Phi_{m n}(\bD\btheta_{1:n}; \bLambda)$ with $\bD$ and $\bLambda$  as in Sect.~\ref{subsec:Smoothing}. Combining these results and recalling the proof of Lemma~\ref{lem1}, if follows that $p( \btheta_{1:n} \mid \by_{1:n} ) \propto \phi_{pn}(\btheta_{1:n}- \bxi; \bOmega)\Phi_{mn}(\bD\btheta_{1:n}; \bLambda)$, which coincides with the kernel of the \textsc{sun}   in Theorem~\ref{thm:JointSmoothing}. \qquad \quad  \qquad  \qquad  \qquad    $\qed$

\vspace{7pt}

\noindent {\bf \em Proof of Corollary \ref{cor:MarginalLikelihood}}. The expression for the marginal likelihood follows  by noting that  $p(\by_{1:n})$ is  the normalizing constant of the smoothing density. Indeed, $p(\by_{1:n})=\int p(\by_{1:n}{\mid} \btheta_{1:n}) p(\btheta_{1:n}) d\btheta_{1:n}$. Hence, the integrand  coincides with the kernel of the smoothing density, so that the whole integral is equal to $\Phi_{m n}(\bgamma_{1:n\mid n};\bGamma_{1:n\mid n})$.  \qquad  \qquad  \qquad   $\qed$

\vspace{7pt}

\noindent {\bf \em Proof of Corollary  \ref{prop1}}. The proof of Corollary \ref{prop1} is similar to that of Lemma \ref{lem1}. Indeed, the proposal $p(\btheta_t \mid \btheta_{t-1}, \by_{t})$ is  proportional to the product between the likelihood $p(\by_{t} \mid \btheta_{t})= \Phi_m(\bB_t\bF_t\btheta_{t}; \bB_t\bV_t\bB_t)$ and  the prior $p(\btheta_t \mid \btheta_{t-1})=\phi_p(\btheta_t-\bG_t\btheta_{t-1};\bW_t)$. To derive the importance weights in \eqref{eq14}, it suffices to notice that the marginal likelihood $p(\by_{t} \mid \btheta_{t-1})$ coincides with the normalizing constant of the \textsc{sun}  in \eqref{eq13}. \qquad \qquad \qquad  \quad  $\qed$

\vspace{7pt}

\noindent {\bf \em Proof of Proposition \ref{prop2}}.
To derive the form of the proposal, first notice that 
$p(\bz_{t-k:t}\mid \bz_{1:t-k-1}, \by_{t-k:t}) \propto p(\bz_{t-k:t} \mid  \bz_{1:t-k-1} ) p(\by_{t-k:t} \mid \bz_{1:t})$. Recalling  model \eqref{eq3}--\eqref{eq5} and Sect.~\ref{sec.422}, we have that $(\bz_{t-k:t}\mid \bz_{1:t-k-1}) \sim \mbox{N}_{m(k+1)}(\br_{t-k:t \mid t-k-1},  \bS_{t-k:t \mid t-k-1})$ and $p(\by_{t-k:t} {\mid} \bz_{1:t}) = \mathbbm{1}(\bz_{t-k:t} \in \mathbb{A}_{\by_{t-k:t}})$. Hence, $p(\bz_{t-k:t} \mid  \bz_{1:t-k-1} ) p(\by_{t-k:t} \mid \bz_{1:t})$ is the kernel of  the $[m(k+1)]$-variate truncated normal in Proposition \ref{prop2}. The form of the weights in \eqref{eq18} follows from their general expression \citep[e.g.,][Sect.~2.2.1]{andrieu2002}, combined with the sequential formulation of the model. Note also that, when written as a function of $\bz_s$ from the proposal, $p(\by_s\mid \bz_s) =1$, for any $s=1, \ldots, t-k$. Therefore, with the convention that $p(\bz_{1}\mid \bz_{0}) = p(\bz_{1})$, the weights are proportional to
\begin{align*}
&\dfrac{p(\bz_{1:t-k}\mid \by_{1:t})}{p(\bz_{1:t-k-1}\mid \by_{1:t-1}) p(\bz_{t-k}\mid \bz_{1:t-k-1}, \by_{t-k:t})}\\
& \propto \dfrac{p(\by_{1:t} \mid \bz_{1:t-k})p(\bz_{1:t-k})/p(\bz_{1:t-k-1})}{p(\by_{1:t-1} \mid \bz_{1:t-k-1}) p(\bz_{t-k}\mid \bz_{1:t-k-1}, \by_{t-k:t})}\\
&\quad = \dfrac{p(\by_{1:t} \mid \bz_{1:t-k})p(\bz_{t-k} \mid \bz_{1:t-k-1})}{p(\by_{1:t-1} \mid \bz_{1:t-k-1})p(\bz_{t-k}\mid \bz_{1:t-k-1}, \by_{t-k:t})}\\
& \quad = \dfrac{ p(\by_{1:t} \mid \bz_{1:t-k})p(\by_{t-k:t} \mid \bz_{1:t-k-1})}{p(\by_{1:t-1} \mid \bz_{1:t-k-1})p(\by_{t-k:t}\mid \bz_{1:t-k})}\\
& \quad =\dfrac{p(\by_{t-k:t}\mid\bz_{1:t-k-1})}
{p(\by_{1:t-1}\mid\bz_{1:t-k-1})}=\dfrac{p(\by_{t-k:t}\mid\bz_{1:t-k-1})}
{p(\by_{t-k:t-1}\mid\bz_{1:t-k-1})},
\end{align*}
where the last equality follows from the fact that $p(\by_{1:t} \mid \bz_{1:t-k})=p(\by_{t-k:t} \mid \bz_{1:t-k})$. To obtain the final form of equation \eqref{eq18} if suffices to note that $p(\by_{t-k:t}\mid\bz_{1:t-k-1}) = \mbox{pr}(\bB_{t-k:t}\tilde{\bz}>{\bf 0})= \Phi_{m (k+1)}(\bmu_t;\bSigma_t)$, where $\tilde{\bz}$ is distributed as a $\mbox{N}_{m(k+1)}(\br_{t-k:t\mid t-k-1},\bS_{t-k:t\mid t-k-1})$, with  $\br_{t-k:t\mid t-k-1}$, $\bS_{t-k:t\mid t-k-1}$, and $\bB_{t-k:t}$ as  in Sect.~\ref{sec.422}. A similar argument holds for the denominator of  \eqref{eq18}.
\qquad $\qed$

\section*{Appendix B: Derivation of computational costs}
\label{sec:compCost}
In this section we derive the computational costs of the algorithms discussed in Sects.~\ref{sec.4} and \ref{sec.5}. Let us first consider Algorithm \ref{algo1} with an initial focus on the smoothing distribution. For this routine, the matrix computations to obtain the parameters of interest require $\mathcal{O}(n^3[p^3+m^3])$ operations. Regarding the sampling cost to obtain  $R$ draws, step [1] requires $\mathcal{O}(p^3n^3 + Rp^2n^2)$ operations since we have to first compute the Cholesky decomposition of $\bar{\bOmega}_{1:n\mid n}- \bDelta_{1:n\mid n}\bGamma_{1:n\mid n}^{-1}\bDelta_{1:n\mid n}^{\intercal}$ in $\mathcal{O}(p^3n^3)$, and then multiply each independent sample for the resulting lower triangular matrix, at   $\mathcal{O}(Rp^2n^2)$ total cost. Step [2] requires, instead, to obtain a minimax exponentially-tilted estimate  at $\mathcal{O}(m^3n^3)$ cost \citep{botev_2017} and then perform $\mathcal{O}(n^2m^2C(mn))$ operations for each independent sample, where $C(d)$ denotes the average number of proposed draws required per accepted sample in \citet{botev_2017}, when the dimension of the truncated normal is $d$.  Hence, the overall cost of Algorithm~\ref{algo1} is $\mathcal{O}(n^3(p^3+m^3) + Rn^2[p^2+m^2C(mn)])$. If the interest is in the filtering distribution, which coincides with the marginal smoothing at $n=t$, it is sufficient to sample $\bU_{0 \ n\mid n}$ instead of $\bU_{0 \ 1:n\mid n}$. Hence, the overall cost for $R$ samples reduces to $\mathcal{O}(tp^3+t^3m^3 + R[p^2 + t^2m^2C(mt)])$.

We now consider the computational costs of the particle filters considered in Sect.~\ref{sec.4} and \ref{sec.5}. For each $t$, the cost is due to computation of parameters, sampling and evaluation of the importance weights. Starting with the ``optimal'' particle filter in Sect.~\ref{sec.421}, the matrix operations for computing the quantities in steps [3.1]--[3.3] of Algorithm \ref{algo2} have an overall cost for the $R$ samples of $\mathcal{O}(m^3 + pm^2 + p^2m + Rpm + Rp^2)$. The sampling costs are, instead, $\mathcal{O}(p^3+Rp^2)$ and $\mathcal{O}(m^3 + Rm^2C(m))$ for the Gaussian and truncated normal terms, respectively. To conclude the derivation of the computational costs, it is necessary to derive those associated with the evaluation of the importance weights. For all the particle filters analyzed, such weights are obtained by evaluating  in $R$ different points the cumulative distribution function of a zero mean multivariate normal with fixed covariance matrix. To facilitate comparison, we assume that this evaluation relies on a Monte Carlo estimate based on $M$ samples in all the particle filters. For the ``optimal'' particle filter, this step requires $\mathcal{O}(m^3 + M m^2)$ operations to obtain the samples, plus $\mathcal{O}(MRm)$ for computing the Monte Carlo estimate. Combining these results, the overall cost for the ``optimal'' particle filter at time $t$ is $\mathcal{O}(t(p^3+m^3) +tR[p^2 + pm + m^2C(m)] + tM[m^2 + Rm])$.

Let us now derive the cost of the Rao--Blackwellized algorithm by \citet{andrieu2002}.
In this case, adapting the notation of the original paper to the one of Sect.~\ref{sec.422}, it can be noticed that one \textsc{kf} step requires $\mathcal{O}(p^3 + Rp^2 + Rpm + m^3)$ operations for the computation of $\bP_{t\mid t-1}, \ba_{t\mid t-1}, \bS_{t\mid t-1}, \br_{t \mid t-1}$, $\bP_{t\mid t}$ and $\ba_{t\mid t}$, at any $t$. As for the sampling part, it first requires $R$ draws from an  $m$-variate truncated normal. Exploiting the same arguments considered for the previous algorithms, this step has an $\mathcal{O}(m^3 + R m^2 C(m))$ cost. The sampling from the final Gaussian filtering distribution $p(\btheta_t\mid \bz_{1:t} = \bz_{1:t\mid t})$ of direct interest requires instead $\mathcal{O}(p^3 + Rp^2)$ operations. Leveraging again the derivations for the previous algorithms, the computation of the importance weights has cost $\mathcal{O}(m^3 + Mm^2 +RMm)$. Therefore, the overall cost of the sequential filtering procedure at time $t$ is $\mathcal{O}(t(p^3 + m^3) + tR[p^2 + pm +m^2 C(m)] + tM[m^2 + Rm])$.

The above derivations for the Rao--Blackwellized algorithm directly extend to the partially collapsed lookahead particle filter shown in Algorithm~\ref{algo4}. In fact, while at each   $t$ the Rao--Blackwellized solution requires one  \textsc{kf} recursion combined with sampling from $m$-variate truncated normals and evaluation of cumulative distribution functions of $m$-variate Gaussians, the  lookahead routine relies on samples from $[m(k{+}1)]$-variate truncated normals along with $k{+}1$ \textsc{kf} steps, and computation of  cumulative distribution functions for $[m(k{+}1)]$-dimensional Gaussians. Hence, adapting the cost of the Rao--Blackwellized algorithm to this broader setting, we have that the overall cost of Algorithm \ref{algo4} at time $t$ is $\mathcal{O}(t(k_{+}p^3 + k_{+}^3m^3) + tR[k_{+}p^2 + k_{+}pm + k_{+}^2m^2C(k_{+}m)]+ tM[k_{+}^2m^2 + Rk_{+}m])$, where $k_{+}=k+1$. Note that, in practice, $k$ is set equal to a pre-specified small constant and, therefore, the actual implementation cost reduces to $\mathcal{O}(t(p^3 + m^3) + tR[p^2 + pm +m^2 C(k_+m)] + tM[m^2 + Rm])$, where $k_{+}$ only enters in $C(k_+m)$.

The bootstrap particle filter leverages the proposal $p(\btheta_t\mid \btheta_{t-1})$, with importance weights given by the likelihood in  equation \eqref{eq1}. Hence, exploiting  similar arguments considered for the previous routines yields a cost $\mathcal{O}(t(p^3+m^3) + tR(p^2+pm) +tM[m^2 + Rm])$.

Finally, note that the cost of the extended Kalman filter \citep{uhlmann1992} is lower than the one of the particle filters since no sampling is involved, except for the Monte Carlo evaluation of the multivariate probit likelihood. In particular, at each $t$, one has to invert a $p\times p$ and an $m\times m$ matrix, plus computing the likelihood, which yields a total cost at $t$ of $\mathcal{O}(t[p^3+m^3 + Mm^2])$.

\end{document}